\documentclass[review,10pt]{elsarticle}

\usepackage{tikz}
\usepackage{amsmath,amsfonts}
\usepackage{subcaption}
\usepackage[inline,shortlabels]{enumitem}
\usepackage{import}
\usepackage{graphicx}
\usepackage{enumitem}
\usepackage{multirow}
\usepackage{xspace}
\usepackage{etoolbox}
\usepackage{verbatim}
\usepackage{setspace}
\usepackage{url}
\usepackage{array}
\usepackage{mathtools}
\usepackage{colortbl}
\usepackage{booktabs}
\usepackage{algorithm}
\usepackage{algorithmic}
\newcommand{\MethodName}{CETRA\xspace}
\makeatletter
\newcommand\newref[1]{#1\def\@currentlabel{#1}}
\newcommand\newtag[2]{(#1)\def\@currentlabel{#1}\label{#2:#1}}
\makeatother

\newcolumntype{x}[1]{>{\centering\arraybackslash\hspace{0pt}}p{#1}}
\newcolumntype{H}{>{\setbox0=\hbox\bgroup}c<{\egroup}@{}}
\begin{document}

%don't want preprint journal printed
\journal{arXiv}
%don't want date printed
%\date{}

\begin{frontmatter}

\title{A Transferable and Automatic Tuning of Deep Reinforcement Learning for Cost Effective Phishing Detection}

\author{Orel Lavie}
\ead{orella@post.bgu.ac.il}

\author{Asaf Shabtai}
\ead{shabtaia@bgu.ac.il}

\author{Gilad Katz}
\ead{giladkz@bgu.ac.il}

\address{
Department of Software and Information Systems Engineering\\
Ben-Gurion University of the Negev\\
Be'er Sheva, Israel
}
\cortext[corref]{Corresponding author.}

\begin{abstract}

%The detection of phishing attacks, like similar challenges in cyber security, requires the deployment of multiple complementary detectors to reach acceptable detection levels. 

Many challenging real-world problems require the deployment of ensembles---multiple complementary learning models---to reach acceptable performance levels. While effective, applying the entire ensemble to every sample is costly and often unnecessary. 
Deep Reinforcement Learning (DRL) offers a cost-effective alternative, where detectors are dynamically chosen based on the output of their predecessors, with their usefulness weighted against their computational cost.
Despite their potential, DRL-based solutions are not widely used in this capacity, partly due to the difficulties in configuring the reward function for each new task, the unpredictable reactions of the DRL agent to changes in the data, and the inability to use common performance metrics (e.g., TPR/FPR) to guide the algorithm's performance. 
In this study we propose methods for fine-tuning and calibrating DRL-based policies so that they can meet multiple performance goals. Moreover, we present a method for transferring effective security policies from one dataset to another. Finally, we demonstrate that our approach is highly robust against adversarial attacks.
\end{abstract}


\end{frontmatter}

\makeatletter
\def\ps@pprintTitle{%
  \let\@oddhead\@empty
  \let\@evenhead\@empty
  \let\@oddfoot\@empty
  \let\@evenfoot\@oddfoot
}
\makeatother

\section{\label{sec:introduction}INTRODUCTION}

%A phishing attack is a common social engineering technique used for tricking victims (end users) into visiting malicious websites and obtaining their sensitive information, which can then be used to gain access to organizations' critical information systems~\cite{chin2018phishlimiter}. 
%Phishing attacks are one of the main threats organizations are facing nowadays, with the cost of successful phishing attacks estimated at 54 million dollars in 2020 alone~\cite{APWG2020,FBI2020}. 
%Phishing attacks are extremely diverse, a fact that has resulted in the development of a variety of detection tools~\cite{mao2013baitalarm,zhang2007cantina,xiang2011cantina+,bahnsen2017classifying,wang2019pdrcnn,mohammadkhani2018new}.

%While these detection tools can provide adequate protection, if maintained and updated, organizations often have to use multiple detectors, and the computational cost of applying many detectors to each web page is prohibitive. 
%Additionally, the amount of time needed to analyze each web page---also a function of the number of detectors---could create a backlog, and hamper an organization's operations. 
%Organizations therefore must strike the right balance between \textit{security}, i.e., protecting their assets, and \textit{availability}, i.e., the ability to provide access in an acceptable time. 

Ensembles are a common approach to improving the performance of machine learning algorithms. Instead of relying on a single algorithm, ensembles combine the output of several learning models to obtain a more accurate result. Ensembles are commonly used not only to improve classification or regression performance \cite{ren2016ensemble,dvornik2019diversity,belitz2021evaluation}, but also to create solutions that are more robust against noise \cite{gupta2019dealing,brajard2020combining} and adversarial examples \cite{li2020adversarial,yang2020dverge}.

While the advantages of using ensembles are considerable, they are not without drawbacks. First, running ensembles can be computationally costly because of the need to train and apply multiple learning models prior to producing a classification. Secondly, while the individual learning models of the ensemble can usually be run in parallel, they all need to conclude their processing of the analyzed data before the ensemble can produce its output. These two drawbacks are particularly challenging in domains where ensembles consist both of a large number of learning models, with large variance in the running times of the individual models. In such cases, the processing of an item may be delayed for a long period of time due to a single model out of dozens.

The aforementioned drawbacks make clear the need for a more refined solution, that can weigh the benefit of using each individual learning model in an ensemble against its cost (measured by running time, computing cost, etc.). Such an elegant solution to this problem was recently proposed in~\cite{birman2022cost}, where the authors presented SPIREL, a deep reinforcement learning (DRL) framework for the dynamic sequential allocation of detectors for each sample. 
Based on the scores assigned by previous detectors, SPIREL either allocates additional detectors or produces a classification. 
SPIREL's reward function assigns a value to correct/incorrect classifications, as well as to the runtime required to analyze each sample.
SPIREL was evaluated on malware detection in executable and Android files, and was able to reduce running time by 80\% while reducing the accuracy and F1 metrics by just 0.5\%.

While effective, SPIREL suffers from two limitations that make it difficult to apply in real world scenarios. 
First, \textit{SPIREL lacks the (basic) capability of supervised learning algorithms to operate on predefined true positive (TP)/false positive (FP) goals}. 
This ability is critical, as organizations often define these thresholds so that their day-to-day operations are not disturbed.
In standard supervised learning solutions, organizations that experience too many false alarms would simply increase the confidence score threshold required to identify a sample as malicious, without re-training their models. 
Doing the same for SPIREL, however, would require modifying the reward function and retraining the model.

SPIREL's second limitation, which exacerbates the first, is the \textit{difficulty of achieving and maintaining the desired levels of TP/FP rates}. 
The DRL-agent's reward function, which guides its decision making process, assigns positive and negative rewards for correct and incorrect classifications, respectively. 
The size of these rewards determines SPIREL's policy, 
However, this reward structure is a blunt instrument, and small changes to the reward structure can have a non-linear impact on an agent's behavior. 
Therefore, fine-tuning a DRL agent for specific desired detection levels is difficult and requires multiple runs and delicate optimization. 
Even worse, changes in the data---a common phenomenon in real life---would require performing the tuning process yet again.

In this study we propose \MethodName, a novel approach for dynamically adapting DRL-based methods to achieve and maintain desired levels of performance. 
\MethodName enables organizations to first define desired values for their key performance indicators (e.g., AUC, RAM usage, running time), and then dynamically modifies its reward function to meet these goals. 
This capability alleviates one of the main difficulties in applying DRL to real-world applications: the configuration of the model. 
To further enhance \MethodName's ability to adapt to organizational policy, we present a probability density-based measure that enables us to translate a successful configuration from one domain to another. Finally, we demonstrate that our approach is highly robust against adversarial attacks.

We evaluate \MethodName on two phishing detection datasets, and demonstrate \MethodName ability to offer a highly cost-effective solution: our derived security policy reduces the processing time by as much as 76\% with a negative impact of only 0.25\%-0.35\% on the F1 metric. 
Additionally, we demonstrate how a simple definition of goals for metric values can automatically modify the derived security policy to achieve these desired goals, and we demonstrate how our proposed density-based measure can successfully transplant an effective security policy from another domain.
Finally, we evaluate the robustness of \MethodName to both evasion and resource utilization adversarial machine learning attacks.

To summarize, our contributions are as follows:
\begin{itemize}
    \item We propose a novel approach for configuring the reward function of DRL-based models.
    Our approach enables us to define desired values for various metrics---TPR/FPR, RAM usage, etc.---and have the DRL model automatically adapt its behavior in order to reach them. 
    
    \item We propose a process for ``transferring'' effective policies from one domain to another. \MethodName is therefore able to automatically and efficiently configure our DRL-based approach.
   
    \item Finally, we demonstrate that our proposed approach is highly robust against adaptive adversarial attacks. 
\end{itemize}

\section{\label{sec:relatedworks}RELATED WORK}

\subsection{Phishing Detection Methods}

Earlier works in phishing detection involved the use of blacklisting---a repository of web pages known to be malicious. 
Such solutions include Google's Safe Browsing and PhishTank services. 
The main weakness of this approach lies in its inability to defend against unknown websites. 
As a result, machine learning is the primary tool used nowadays to fight phishing.

Machine learning-based approaches extract various features from the content of the web page or its metadata to classify a given page as benign or phishing. 
The solutions in this field are numerous and diverse~\cite{xiang2011cantina+,horng2011efficient,zhang2014domain,mohammad2014predicting,verma2015character,moghimi2016new}.
Zhang et al.~\cite{zhang2017two}, for example, used an extreme learning machine (ELM) technique and extracted hybrid features from the page's URL and text. 
Recently, due to their ability to process large amounts of data and learn complex patterns, deep learning (DL) models were proposed for the detection of phishing websites. 
These methods mainly analyze the URL string using LSTM~\cite{bahnsen2017classifying} or CNN~\cite{saxe2017expose,le2018urlnet,wang2019pdrcnn} models.

\subsection{Reinforcement Learning}
Reinforcement learning (RL) is an area of ML that addresses complex multi-step decision-making processes. 
These algorithms have been shown to perform well even with noisy and partial information~\cite{kalweit2017uncertainty}.
An RL algorithm normally consists of an agent that interacts with an environment in a sequence of actions and rewards. 
In each time step $t$, the agent selects an action $a_i \in A$ = $\{a_{1}, a_{2}, ..., a_{k}\}$. As a result of $a_i$, the agent transitions from the current state $s_t$ to a new state $s_{t+1}$. 
Additionally, the selection of the action may yield a reward $r_t$, which can be either positive or negative. 
The goal of the agent in each state is to interact with the environment in a way that maximizes the sum of future rewards $R_{T}$:
\begin{equation}
R_{T} =
\text{\begin{math}  \sum_{t=1}^T r_{t} \end{math}}
\end{equation}
where $T$ is the terminal (final) step in the sequence.

The selection of actions is made by the policy $\pi(a,s)$, which produces the probability of taking action $a\in A$ when in state $s\in S$. 
A common approach for evaluating the future rewards of an action is by its Q-function, denoted by $Q(s,a)$, which calculates the expected reward $E[R_t|s_t=s,a_t=a,\pi]$, derived from the pair $s$,$a$ for policy $\pi$ at time $t$. 
In problems consisting of very large state and/or action spaces, estimating $Q(s,a)$ for every possible state-action combination is infeasible. Therefore, it is common to use an approximation-based solution, such as a neural network. 
Instead of $Q(s,a)$, we use $Q(s,a;\theta)$, where $\theta$ reflects the parameters of the neural network, making it a deep reinforcement learning (DRL) setup~\cite{van2016deep,mnih2015human}.

\begin{comment}

Fundamental concepts in reinforcement learning are defined as follows:
\begin{itemize}
    \item $States.$ The set of states $S$, also called the environment, representing all possible scenarios (i.e., continuous combinations) the agent may encounter.

    \item $Actions.$ The set of actions $A$ representing all possible steps taken by the agent.

    \item $Rewards.$ The rewards and penalties (i.e., "costs") R(a,s) assigned to an action $a\in A$ performed in state $s\in S$.

    \item $Policy.$ The policy $\pi(a,s)$ maps the probability of taking action $a\in A$ when in state $s\in S$; for input state $s$ and action $a$, the policy $\pi(a,s) = P(a_t = a | s_t= s)$ at time $t$.

    \item $Q\_function.$ The $Q(s,a)$ for input state $s \in S$ and action $a \in A$ is a common method used to calculate the expected reward $E[R_t|s_t=s,a_t=a,\pi]$ derived from the pair $s$,$a$ for policy $\pi$ at time $t$ \cite{sutton2018reinforcement,fan2020theoretical}.
\end{itemize}
\end{comment}

DRL-based algorithms have several properties that make them highly suitable for the security domain. 
First, they have the capability of efficiently exploring large state and action spaces~\cite{colas2018gep}, and of devising strategies to address complex problems \cite{zhou2018deep}. 
Additional properties that make DRL useful in security-related scenarios is their ability to operate with partial~\cite{kalweit2017uncertainty} and noisy information~\cite{feng2018reinforcement}. Finally, their use of the reward function to shape the policy of the algorithm enables the reconciliation of multiple (sometimes conflicting) objectives~\cite{birman2022cost}.

%The ability of DRL algorithms to explore large solution spaces, and create highly efficient strategies to address them, has proven highly effective in various areas such as robotics and control problems~\cite{schulman2015trust}. 
%In addition, DRL performs well in scenarios of high uncertainty and partial information~\cite{kalweit2017uncertainty}, and in cases where efficient exploration/exploitation strategies are crucial for reaching a solution~\cite{ishii2002control}. 

\subsection{Reinforcement Learning-Based Security}
The sequential decision making employed by DRL algorithms make them highly suitable for the development of security policies. 
%However, the difficulties in configuring and modifying said policies have had a chilling effect on its application in security-related fields. \gilad{I'm not sure I like this claim. It's a little risky to say something like this - how many studies are there in this field?} \orel{I guess you are right, there are fewer works in RL, but not as much as it sounds from the sentence. However, I think that this sentence refers to security ensembles deployed via RL frameworks specifically, which are indeed fewer.}
%\orel{How about a comparison with security in general deep learning? or comparing it to the security ensemble case? it sure has much less articles}.
Blount et al.~\cite{blount2011adaptive} presented a proof of concept for an adaptive rule-based malware detection framework for portable executable (PE) files that employs a classifier combined with a rule-based expert system. 
Then, an RL algorithm is used to determine whether a PE is malicious. 
%Sepideh et al.~\cite{mohammadkhani2018new} proposed using RL to classify different malware types, utilizing a set of features commonly used by antivirus software.
Birman at el.~\cite{birman2022cost} proposed a DRL framework for the dynamic sequential allocation of detectors for each sample.
Based on the scores assigned by previous detectors, their framework either allocates additional detectors or produces a classification. 
%The reward function of this approach assigns a value to correct/incorrect classifications, as well as to the runtime required to analyze each sample.
In the field of email phishing detection, Smadi et al.~\cite{smadi2018detection} proposed a two-mode framework.
In the online mode, the framework employs a DRL agent to detect phishing attacks. 
Offline, the model adapts to changing circumstances by performing additional learning. 
%SPIREL was evaluated on PE files and Android files, and was able to reduce running time by 80\% with minimal impact on detection rates.

Fu et al.~\cite{fu2017learning} introduced adversarial inverse reinforcement learning (AIRL), a practical and scalable inverse RL algorithm based on an adversarial reward learning formulation. 
Another work in the field of adversarial learning by Anderson et al.~\cite{anderson2018learning} presented a reinforcement learning method that learns which sequences of operations are likely to enable a sample to avoid detection. 
Recently, Mo et al.~\cite{mo2022attacking} introduced Decoupled Adversarial Policy, which uses two sub-policies: one that determines when to launch an attack, and another that determines what actions the model should be ``lured'' into taking.

\begin{comment}
In the field of adversarial learning, Fu et al.~\cite{fu2017learning} introduced adversarial inverse reinforcement learning (AIRL), a practical and scalable inverse RL algorithm based on an adversarial reward learning formulation. 
The study demonstrated AIRL's ability to recover reward functions that are robust to changes in dynamics, enabling it to learn policies even when there is significant variation seen in the environment during the training. 
Their experiments showed that AIRL outperforms existing methods in these settings. 
Anderson et al.~\cite{anderson2018learning} presented a reinforcement learning method that learns which sequences of operations are likely to evade through a series of games played
against a gradient-boosted based anti-malware engine.
Their study also showed that attacks against the gradient-boosted model appear to evade components of publicly hosted antivirus engines.
In the field of email phishing detection, Smadi et al.~\cite{smadi2018detection} proposed a framework that has two modes: online and offline. 
In the online mode, that framework combines a neural network with RL to detect phishing attacks. 
To reflect changes in newly explored behaviors, their model can adapt itself to produce a new phishing email detection system when it is needed. 
The proposed model solves the problem of a sparse dataset by automatically adding more emails to the offline dataset in the online mode.
\end{comment}
%In Table ~\ref{tbls:tbl_related_works} we summarize the methods mentioned above.

\section{PROPOSED METHOD}
\label{sec:method}

\MethodName builds upon the work of Birman et al.~\cite{birman2022cost}, which proposed SPIREL, a DRL-based approach for the efficient utilization of ensembles: rather than deploy all detectors at once, the agent dynamically select which additional detectors (if any) to call based on the results of previous ones. While highly effective, SPIREL is hindered by its inability to adapt its policy to achieve specific performance metric goals (e.g., false-positive rate of no more than 1\%), or to easily keep these metrics stable in the face of changing data. In this section we present our proposed solution to these limitations.

This section consists of two parts: in Section~\ref{subsec:baseMethod}, we present the basic building blocks of our approach: states, actions, and preliminary reward function. 
In Section~\ref{subsec:Method} we present our proposed expansions, which address the multiple limitations that complicate the deployment of DRL-based solutions to real-world scenarios.

%In this section we present \MethodName, a DRL-based approach for cost-effective phishing detection. The goal of our proposed approach is to devise a solution for the \textit{dynamic application of detectors for cost-effective classification}. Rather than apply all available detectors for each sample, our goal is to deploy the minimal number of detectors required to correctly classify a sample, thus conserving the resources that would otherwise be ``wasted''. While the idea itself has been proposed in the recent work of \cite{birman2022cost}, their approach is hindered by the shortcomings described in Section \ref{sec:introduction}. In this section we describe our proposed approach for addressing these shortcomings.

\subsection{Base Method}
\label{subsec:baseMethod}

In this section we present the states, actions, and rewards representation of our proposed approach. 
Our representation closely follows that of~\cite{birman2022cost}, as we build upon this base in the following section.\\

%In this section we present the building blocks of our proposed approach: states and actions space representations, as well as the reward function. Our representation closely follows that of~\cite{birman2022cost}, who originally proposed the idea of cost-effective ensemble classification. In addition to providing a solid foundation for our proposed novelties (i.e., expansion of this base model), this setting enables us to compare ourselves to the current state-of-the-art solution. 
%We now describe the three components of our DRL-based approach: the states, actions, and the reward function.\\

\noindent \textbf{States.} Our state space consists of all possible combinations of detector outputs. 
Therefore, for a given set of detectors $D$, we represent each state $s \in S$ using a vector of size $|s| = |D|$.
Each entry in the vector represents the output produced by its corresponding detector, which can be considered as the certainty level---the degree to which the detector is certain of the web page's maliciousness.
For yet-to-be-activated detectors we use a default value of -1: 

\begin{equation}
	v_{i} =
	\begin{cases}
		[0,1] & ,\text{detector prediction, if used}\\
		-1 & ,\text{otherwise}
	\end{cases}  
\end{equation}

\noindent \textbf{Actions.} Our action space consists of two types of actions: \textit{detector activation} and \textit{classification}. 
The former is used to apply one detector on the analyzed web page, so the number of such actions is equal to the number of detectors $|D|$. 
The latter action type is used to issue a final classification for the analyzed web page: either ``phishing'' or ``benign''. 
Applying either of their final classifications also terminates the analysis of the web page.\\

\noindent \textbf{Rewards.} The reward function shapes the DRL agent's policy. Once our approach classifies a web page, there are four possible outcomes: true positive/negative (TP and TN) and false positive/negative (FP and FN). The \textit{base reward function} used by our approach (as well as SPIREL) is as follows:
\begin{equation}
C_{1_{total}}(T) =
    \begin{cases}
    r & ,\text{TP or TN}\\
    -1*\sum_{t=1}^{T} C_{1}(t) & ,\text{FP or FN}\\
    \end{cases}
\label{funcs:ref_func}
\end{equation}

where r is a constant and $C_1(t)$ is a time-dependent loss function that the ``punishment'' it assigns to the DRL is proportional to the amount of computing resources spent on the classification:

\begin{equation}
C_{1}(t) =
    \begin{cases}
    t & ,\text{$0 \leq t < 1 $}\\
      1+\log_2 \left( \text{min} \left( t,t_1(s)\right) \right) &   ,\text{$ 1 \leq t $}\\      
     
    \end{cases}
\label{funcs:ref_func_positive}
\end{equation}

The rationale of the approach is straightforward: a fixed reward for correct classifications means that detectors that can provide additional information (i.e., increase certainty) are likely to be used. 
However, the risk of being mistaken will prevent the DRL-agent from calling upon detectors that provide little to no information, because their use will only incur a larger loss in case of a mistake.\\

\noindent \textbf{Limitations of the base method.} While this approach performed well in simulation \cite{birman2022cost}, we identify two significant shortcomings to this approach when applied to real-world use-cases:
\begin{enumerate}
    \item \textbf{Lack of theoretical or practical guidelines} -- SPIREL does not offer any method for configuring the $r$ and $C_1(t)$. This makes the approach difficult and computationally costly to implement for new datasets.
    \item \textbf{The need to conform with multiple objectives} -- organizations often define multiple metrics for detection frameworks (e.g., TPR and running time). Manually configuring the reward function---the only level offered by SPIREL---to meet these goals can be computationally prohibitive because of the large search space.
\end{enumerate}

\noindent In the next section we present our proposed approach for addressing these challenges.

\subsection{\MethodName}
\label{subsec:Method}

This section comprises three parts.
In section~\ref{subsubsec:TheoryAnalysis} we develop a theoretic foundation for the required traits of an effective multi-objective reward function. 
In Section~\ref{subsubsec:MetricCostFunction}, we present our novel approach for reformulating the reward function in a way that enables our DRL-agent to automatically adapt its behavior to comply with multiple objectives without any need for additional calibration. Finally, in Section \ref{subsubsec:KLbasedPolicyTransfer} we present a novel method for ``importing'' the settings of one successfully deployed DRL-agent to another.

\subsubsection{Theoretical analysis of the cost function}
\label{subsubsec:TheoryAnalysis}

We now formally define guidelines for the design of the reward function used by our approach. 
Such a definition is important for several reasons. 
First, such guidelines will produce reward functions with stable and predictable outputs that will facilitate the convergence of the DRL model. 
Secondly, our guidelines provide a simple framework for the representation of multiple factors (e.g., performance, runtime, RAM usage) in a single reward function, thus making our approach applicable to multiple use-cases and domains. 
Thirdly, reward functions created using our guidelines can be adapted to new domains, thus preventing exhaustive exploration of the reward function for new domains. 
Our guidelines are as follows:

%\begin{enumerate}
    \noindent \textbf{1) Consistent reward structure} -- ``good'' and ``bad'' outcomes will receive positive and negative values, accordingly. 
    
    \noindent \textbf{2) Weak monotonous function} -- for every outcome, the cost should not decrease as a function of the invested time and/or resources.
    
    \noindent \textbf{3) Bounded on both axes} -- the reward function should be bounded on all axes to prevent outliers and noisy data from having an out-sized effect on the DRL agent.
    %For this reason, the values of the vertical axis should also be bounded, thus placing a bound on the positive and negative rewards that can be obtained for any sample.
    
    \noindent \textbf{4) Continuous and proportional} -- To prevent small changes in the input (e.g., running time) from triggering large fluctuations in the reward function values, we require that the latter be continuous. 
    Moreover, we require the reward function to be a \textit{linear or a super-linear} expression of the input. 
    This setup enables us to increase/decrease the change rate of the reward, thus enabling organizations to define different priorities to different ranges of values (e.g., low/high computational cost for the analysis of a sample). 
    While different priorities can be defined, the linear change rate keeps the change gradual and predictable.

\subsubsection{Metric Goal-driven Cost Functions}
\label{subsubsec:MetricCostFunction}

As explained in Section~\ref{sec:introduction}, the two main challenges not addressed in SPIREL were \textit{a)} the need to automatically adapt the reward function to obtain the organization's detection objectives (e.g., TPR/FPR), and; b) to smoothly adapt the reward function to changes in the dataset's characteristics over time, so that the aforementioned goals are maintained. We now describe three modifications to the base approach, designed to overcome these challenges.

\noindent \textbf{Step 1: reformulation of the different regions, defined by the reward function}. 
Let $PR(a \leq x \leq b)$ be the probability density function for the computational cost of the processing samples from a given dataset $D$. 
We define the computational cost of a sample as the cost of analyzing it using all its assigned detectors. 
The cost can be defined by the organization, and include multiple factors---running time, cloud credits usage, etc.---but in order to make our results comparable to previous work~\cite{birman2022cost}, we use running time as a metric. By considering running time as our cost approximator, the probability density function enables us to obtain the percentage of samples whose processing time is between $[a,b]$.

Our proposed representation has two significant advantages.
First, it enables the creation of dynamic boundaries that can be automatically adapted to changes in the dataset over time. 
We are therefore able to define different cost functions to different percentiles of samples, something that was not possible while using the rigid boundaries proposed by~\cite{birman2022cost}. 
Secondly, this representation provides the theoretical foundation that enables the ``translation'' of successful policies from one domain to another (see Section \ref{subsubsec:KLbasedPolicyTransfer}).

Without loss of generality, we assume our reward function to have two `regions', each with a different reward slope. The new function (Equation \ref{funcs:our_func_positive}) will now form the base for our additional modifications.

\begin{equation}
C_{2_{total}}(T)
    \begin{cases}
    r & ,\text{TP or TN}\\
    -1*\sum_{t=1}^{T} C_{2}(t) & ,\text{FP or FN}\\
    \end{cases}
\label{funcs:our_func_main}
\end{equation}

where $r$ is a constant and $C_2(t)$ is defined as follows:
\begin{equation}
C_{2}(t) =
    \begin{cases}
      \frac{t}{d_2} & ,\text{$0 \leq t < d_2 $}\\
      1+\log_2 \left( \frac{\text{min} \left( t,t_2(s)\right)}{d_2} \right) &   ,\text{$ d_2 \leq t $}\\
    \end{cases}
\label{funcs:our_func_positive}
\end{equation}

\noindent \textbf{Step 2: incorporating organizational priorities into the reward function.} As explained earlier, one of SPIREL's main shortcomings is its inability to incorporate metrics---accuracy, TPR/FPR etc.---into the reward function. We now present a new reward function that achieves this goal:

%The next modification we made to the reward function relates to the DRL-agent's compliance with the organization's desired goals. 
%While the reward function defined in equation~\ref{funcs:our_func_main} is a function of consumed resources (which is desirable), it does not consider whether the organization's desired metric goals, be they TPR/FPR, accuracy or any resource-based metric (which is undesirable). 
%To induce the DRL-agent to take these goals into account, we create a new reward function that builds upon the previous one:

\begin{equation}
C_{2_{metric}}(m) =
    \begin{cases}
      -b * CR(M) & ,\text{$m < l$}\\
      CR(M) & ,\text{$l \leq m < u$}\\
      b * CR(M) & ,\text{$u \leq m$}\\
    \end{cases}
\label{funcs:extension_func_main_T}
\end{equation}
\noindent where $u$ and $l$ are the upper and lower bounds of the acceptable range of the evaluated metric, and $m$ is the value obtained by our approach for a batch of $M$ samples. The parameter $b$ is a manually defined bonus/penalty factor, and $CR(M)$ is defined as follows:

\begin{equation}
CR(M) =
    \sum_{i=1}^M C_{2_{total}}(T_i)
\label{funcs:our_func_total_cur}
\end{equation}
\noindent where $\{T_i\}_{i=1}^M$ are the terminal steps for the last $M$ samples.

The rationale of equation \ref{funcs:extension_func_main_T} is as follows: if the reward function $C_{2_{total}}$ yields results that are within the organization's desired range, the reward is unchanged.
If, however, the results are below the predefined range (i.e., $m < l$), then $CR(M)$ is significantly reduced. 
On the other hand, performance that is above the desired range will be rewarded by a bonus. 
This reward setting is designed so that the DRL-agent's first priority will be meeting or surpassing the organization's predefined goal, since achieving this goal is crucial for obtaining high rewards.
Once this goal has been achieved, the DRL-agent will turn its attention to optimizing $CR(M)$, since any increase in the output of this function will result in a linear or super-linear increase in the value of $C_{2_{metric}}$.

\noindent \textbf{Step 3: Measuring organizational metrics over batches.} While logical, a simplistic implementation of the modification proposed in step 2 is not practical. 
The reason is that for many metrics---TPR and FPR are two clear examples---we would need to process all the samples in the training set before calculating $C_{2_{metric}}$ and updating the weights of our neural net. This would significantly slow our model's convergence.

Our third and final modification to the reward function is the calculation of the $C_{2_{metric}}$ reward function over \textit{batches}. 
For each of the analyzed batches of samples in the training phase, we calculate $CR(M)$ and then determine, according to the value of $m$ calculated for the said batch, how $CR(M)$ should be modified. While this approach inevitably injects some volatility into the DRL-agent's learning, this volatility is balanced out over a large number of batches.
%\orel{ Should we mention here the trade off between small batch size and very small one? i.e., large number is avoided to balance the volatility (an epoch equals to batch-size * numb-of-batches). For very small batch sizes could we assume that they are "approaches" to the basic case with batch size of 1 which of course is not mentioned, because it is a special case that a batch is not needed. We should require that $ \frac{1}{m} \leq u-l => m \geq \lceil \frac{1}{u-l} \rceil$, where  $ \frac{1}{m}$ is the "resolution" of the minimum deviation between two metric values, and it actually requires that at least one point should be in the interval [l,u].  For example, in the case of m=10, and l=0.95 and u=0.97, the metric r could get just -bonus or +bonus and not in the middle (could be 0.9 or 1 and not between), because $\frac{1}{10} > 0.02$. } \gilad{don't see the need. overly complex}
As an additional smoothing mechanism, we ensure that the batches are randomly re-sampled at every epoch. As shown by our evaluation, this approach is effective in enabling \MethodName to converge to an effective detection policy.

%\orel{change the order of transfer and metric in the method?}
\subsubsection{Density Function-based Security Policy Transfer}
\label{subsubsec:KLbasedPolicyTransfer}

In the previous section, we described how the reward function of our approach can be designed so that it reflects both the \textit{actual costs of classification} on the one hand, and the \textit{organization's performance goals} on the other. We now seek to make our approach transferable, by enabling the sharing of successful policies across multiple domains.

While their metrics (e.g., TPR/FPR) may vary, organizations are likely to find it more difficult to quantify the cost of a false classification as a function of resource use, as defined in equation \ref{funcs:our_func_main}. 
We therefore propose an approach that enables the transfer of effective settings to a new dataset. 
Please note that while we define our approach as one that operates across datasets, it can also be used to re-calibrate the cost function in cases of data and concept drift.

As defined in Section~\ref{subsubsec:MetricCostFunction}, the boundaries of the cost function are defined using a probability density function. 
Let $C_n$ be the cost function of the dataset $D_n$. That is, we define two instances of equation \ref{funcs:our_func_positive} with specific values of $t$ and $s$ (i.e., a chosen time percentile). Next we define a continuous random variable $X_n$ with the probability density function (i.e., PDF) as:

\begin{equation}
f_{X_n}(t) =
    \begin{cases}
      \frac{C_{n}(t)}{k_n} & ,\text{$0 \leq t \leq t_n(s)$}\\
      0 &   ,\text{otherwise}\\
    \end{cases}
\label{funcs:PDF_Xn_func}
\end{equation}

where $k_n$ is defined as:
\begin{equation}
k_n = \int_{0}^{t_n(s)} C_n(t) dt
\label{funcs:PDF_kn_norm}
\end{equation}

Equation~\ref{funcs:PDF_kn_norm}, enables us to ensure that the properties of $f_{X_n}(t)$ are those of a PDF. 
Next we define a discrete random variable $Y_n$, an indicator function of $X_n$ on a subset $A = [d_n,t_n(s)]$ as:
\begin{equation}
Y_n = \textbf{1}_A (X_n) =
    \begin{cases}
      1 & ,\text{$X_n \in A$}\\
      0 &   ,\text{$X_n \notin A$}\\
    \end{cases}
\label{funcs:indicator_Yn_func}
\end{equation}

This indicator is used to determine whether $X_n$ is below (i.e., $Y_n = 0$) or above (i.e., $Y_n = 1$) the threshold. Because $Y_n$ is a discrete random variable, we can create a vectorized representation of the probability mass function $P_{Y_n}$ of $Y_n$ as:

\begin{equation}
P_{Y_n} = \begin{bmatrix}\mathbb{P}(Y_n = 0), & \mathbb{P}(Y_n = 1)\end{bmatrix}
\label{funcs:PMF_indicator_n}
\end{equation}
where $\mathbb{P}(Y_n = 0)$ and $\mathbb{P}(Y_n = 1)$ are defined as:
\begin{equation}
\mathbb{P}(Y_n = 0) = \int_{0}^{d_n} f_{X_n}(t) dt \quad \mathbb{P}(Y_n = 1) = \int_{d_n}^{t_n(s)} f_{X_n}(t) dt
\label{funcs:pyn1_pyn0}
\end{equation}
Based on equation~\ref{funcs:PMF_indicator_n} and the definition of probability, we derive the following:
\begin{equation}
\mathbb{P}(Y_n = 0) + \mathbb{P}(Y_n = 1) = 1
\label{funcs:sum_pyn1_pyn0_1}
\end{equation}
Based on equation~\ref{funcs:PMF_indicator_n} and equation~\ref{funcs:pyn1_pyn0}, we define a likelihood ratio (i.e., a ratio between the density of lower and higher resources) $\beta_n$ as:
\begin{equation}
\beta_n = \frac{\mathbb{P}(Y_n = 1)}{\mathbb{P}(Y_n = 0)} = \frac{\int_{d_n}^{t_n(s)} f_{X_n}(t)dt}{\int_{0}^{d_n} f_{X_n}(t)dt} = \frac{\int_{d_n}^{t_n(s)} C_n(t)dt}{\int_{0}^{d_n} C_n(t)dt}
\label{integrals:pdf_to_areas_n}
\end{equation}
Based on equations~\ref{funcs:sum_pyn1_pyn0_1} and ~\ref{integrals:pdf_to_areas_n}, we derive the following: 
\begin{equation}
\mathbb{P}(Y_n = 1)=\frac{\beta_n}{\beta_n+1}  \quad \mathbb{P}(Y_n = 0)=\frac{1}{\beta_n+1}
\label{equations:bn_y1_y0}
\end{equation} 
Let $n={n_1}$ and $n={n_2}$ for datasets $D_{n_1}$ and $D_{n_2}$, respectively.
The Kullback–Leibler (i.e., KL) divergence \cite{raiber2017kullback} between $Y_{n_1}$ and $Y_{n_2}$ for ${n_1}\ne {n_2}$ is defined as:
\begin{equation}
D_{KL} \left( Y_{n_1} \| Y_{n_2} \right) = \sum_{y=0}^1 \mathbb{P}(Y_{n_1}=y) \log_{10} \left( \frac{\mathbb{P}(Y_{n_1}=y)}{\mathbb{P}(Y_{n_2}=y)} \right)
\label{equations:D_KL_i_j}
\end{equation}

Using equations~\ref{equations:bn_y1_y0} and \ref{equations:D_KL_i_j}, we derive the following:
\begin{equation}
\begin{array}{l}
D_{KL} \left( Y_{n_1} \| Y_{n_2} \right) = \sum_{y=0}^1 \mathbb{P}(Y_{n_1}=y) \log_{10} \left( \frac{\mathbb{P}(Y_{n_1}=y)}{\mathbb{P}(Y_{n_2}=y)} \right) =  \\
= \log_{10}\left(\frac{\beta_{n_{n_2}}+1}{\beta_{n_1}+1} \right)+\frac{\beta_{n_1}}{\beta_{n_1}+1}\log_{10}\left(\frac{\beta_{n_1}}{\beta_{n_2}} \right)
\end{array}
\label{equations:D_KL_Bi_Bj}
\end{equation}

To require that the density of $D_{n_2}$ is similar to $D_{n_1}$, and since the KL metric is not symmetrical~\cite{tabibian2015speech}, we require that:
\begin{equation}
0 <
\frac{D_{KL} \left( Y_{n_1} \| Y_{n_2} \right)+D_{KL} \left( Y_{n_2} \| Y_{n_1} \right)}{2} < \varepsilon
\label{inequalities:avg_yi_yj}
\end{equation}
where $\varepsilon$ is a hyper-parameter tolerance that decides how much the distributions could differ. 

In a use-case where $C_{n_1}$ and $d_{n_1}$ are known, the above equations enable us to calculate the values of $\beta_{n_2}$ and $d_{n_2}$, and to derive the value of $C_{n_2}$ that will enable us to achieve comparable performance. This holds true even if the sample distribution of the two datasets, as well as their costs, are different.

The method is further discussed over \ref{sec:Appendix_transfer_theorem_algo}, and the full algorithm is also provided there.
%in Appendix~\ref{subsec:Appendix_transfer_algo}.

\section{\label{sec:expersetup}EXPERIMENTAL SETUP}
%This section is organized as follows: in Section~\ref{subsec:datasets} we describe the datasets used in our evaluation and in Section~\ref{subsec:detectors} we describe the detectors that are deployed by our DRL-agent. 
%Finally, in Appendix~\ref{subsec:detectorPerformance} we analyze the performance of our chosen detectors and provide statistics on their running times, performance, and ability to complement each other's detection capabilities.

\subsection{\label{subsec:datasets}Datasets}
Our evaluation was conducted on two datasets, which we name based on the studies that presented them. The first dataset is \textit{Bahnsen} \cite{bahnsen2017classifying}, which contains 1.2M benign URLs taken from the Common Crawl corpus, and 1.146M phishing URLs from PhishTank. The second dataset is \textit{Wang}~\cite{wang2019pdrcnn}, which contains 245K phishing URLs collected from PhishTank, and 245K benign URLs collected by searching via different search engines in domains belonging to Alexa 1M ranking URLs, and returning the URLs of the top ten search results. 

Because we employ a content-based detector in our experiments, we filtered any URL that could not be reached or lost its phishing activity at the time of our evaluation. 
Consequently, the Bahnsen dataset was reduced to 687,176 samples overall---354,770 benign and 332,406 phishing. 
Similarly, the Wang dataset was reduced to 327,646 samples---214,034 benign and 113,612 phishing.

\subsection{\label{subsec:detectors}The Detectors}
Our selection of the detectors was guided by two objectives:

%\begin{itemize}
    \noindent \textbf{1) Variance in resource consumption.} To demonstrate \MethodName's cost-effectiveness in dynamic detectors selection, the detectors need to have different resource consumption (e.g., runtime) costs.
    
    \noindent \textbf{2) Detection methods diversity.} To ensure that our detectors have a diverse set of capabilities (and to better model a real-world detection framework), we chose detectors that use various phishing detection methods (e.g., string based and content based).
    
    %\item Third, the generic solution objective. 
    %The usage of a phishing detection framework without any special configuration or adaptation requirement proves the generalization of the provided solution. \gilad{didn't understand this one, so I didn't change it.}
    %\orel{The main point is that we did not require any special demand besides diverse base detectors (not even specific ones) to make the method work. Thus, it is a general method. That is how I see it.}
%\end{itemize}

Based on the aforementioned objectives, we chose to include \textit{five detectors} in our evaluation, all based on studies from recent years.

\textbf{CURNN}~\cite{bahnsen2017classifying}. This detector directly analyzes the URL's character sequence. 
Each characters sequence exhibits correlations with other characters in the sequence, and its underlying assumption is that these correlations could be leveraged for the identification of phishing. 
We use the implementation described in~\cite{bahnsen2017classifying}, which uses a standard LSTM architecture. 

\textbf{eXpose}~\cite{saxe2017expose}. This detector receives as input multiple unprocessed string-representing URLs, registry keys etc. 
These inputs are projected into an embedding layer and then processed using convolutional layers, and finally fed to a dense layer that classifies each sample using a softmax function as phishing or benign.

\textbf{PDRCNN}~\cite{wang2019pdrcnn}. Encodes the URL as a two-dimensional tensor, and feeds it into a bidirectional LSTM network. 
The LSTM extracts a representation for each character in the URL, which is then provided as input to a convolutional architecture that identifies the key characters required for the detection of phishing.

\textbf{FFNN} This content-based detector utilizes the 17 features presented in~\cite{moghimi2016new} and the 12 features presented in~\cite{jain2019machine}. 
Both studies extract features from the URL string itself and the content of the page (e.g., external and internal hyperlinks ,etc). 
Based on these two studies, we used a dense architecture with ten layers.

\textbf{XGBoost}~\cite{chen2015xgboost}. This popular and highly-effective tree-ensemble algorithm applies both boosting and gradient descent. 
We used five Markov Chain-based models to extract features that are then used as input. 
Our Markovian models were applied on the 1M Alexa rank dataset as follows: Alexa 1M domain uni-gram, Alexa URL uni-gram, Alexa URL parts (i.e., tld, domains and sub-domains), Alexa URL bi-gram and DNS URL bi-gram.

While the selection of diverse detection methods is important for an ensemble solution, diversity alone is not sufficient. A comprehensive analysis is performed in \ref{subsec:detectorPerformance} shows that our chosen detectors are complementary, and that no detector dominates another (i.e., achieves equal or better results for all samples).

\subsection{Experimental Setting}
We ran each detector as an isolated process on a dedicated and identical machine to ensure that the analysis is unbiased.
We used a train/validation/test split of 75\%/10\%/15\%.
Our proposed framework was implemented using Python v3.8 and OpenAI Gym~\cite{brockman2016openai}.
We used the actor-critic with experience replay (ACER)~\cite{lin1992self} (an offline learning DRL algorithm) as our DRL-agent, and used ChainerRL~\cite{nandy2018reinforcement} for the implementation. 
We used an Intel Xeon CPU E5-2690 v4 with 6 cores, 56GB of RAM, and 340GB of SSD disk space.

We set the size of the replay buffer to $5,000$, and begin using it in the training process after $100$ episodes. 
Both the policy and action-value networks of the ACER consist of the following architecture: input layer of size $5$ (the state vector’s size), a hidden layer of size $32$, and an output layer of size $7$ (we have seven possible actions: five individual detector selections, and two final classifications). 
All layers except for the output layer use the ReLU activation function, while the output layer uses softmax.
We set our initial learning rate to $0.001$, with an exponential decay rate of $0.99$ and $\epsilon=1e\textbf{-}8$. The chosen optimizer was RMSPropAsync, which was shown to perform well for DRL tasks~\cite{tieleman2012lecture}. For the overall evaluation, we use the precision and recall measures, as well as the $F1$ measure, which is a combination of the two \cite{zhang2015estimating}.

%In addition to the reward function described in Section~\ref{subsec:Method}, two use-cases that may occur during training immediately incur a significant negative reward and a termination of the current episode (i.e., sample analysis). 
%The two use-cases are the activation of a previously activated detector (which wastes resources and grants no new information), and the attempt to produce a final classification without calling any detectors (an illogical move). 
%Both use-cases incur a negative reward of $-10,000$. 
%As a result, the DRL-agent stops performing these actions relatively early in its training process.

\subsection{The Evaluated Baselines}
We compare \MethodName to two baselines:
%\begin{itemize}
    
    \noindent \textbf{SPIREL~\cite{birman2022cost}.} The original approach for cost-effective ensemble classification, described in Section \ref{subsec:baseMethod}.
    Given that our proposed novelties build upon this approach, a comparison is called-for.
    
    \noindent \textbf{Detector combinations.} Because \MethodName is based on the intelligent and dynamic deployment of detectors, it is important that we show that no static combination of the detectors (or a subset of them) can outperform our proposed approach. 
    For this reason, we evaluate \textit{every possible detector combination}.
    We also use four methods for aggregating the classification results of the evaluated detectors: \textit{a)} majority voting -- the averaging of the confidence scores of all participating detectors; \textit{b)} the ``or'' approach, which returns the maximal score from all applied detectors; \textit{c)} stacking -- we feed all the confidence scores into Random Forest and Decision Tree algorithms, and use them to produce the final confidence scores, and;  \textit{d)} boosting -- we use Adaboost algorithm to produce final scores.
    Overall, we evaluate 61 detector/aggregation configurations for each dataset.

\subsection{The Baseline Reward Functions}
The SPIREL baseline \cite{birman2022cost} was evaluated using five reward function configurations, which were used primarily to demonstrate SPIREL's ability to perform well at various cost/performance trade-offs. Another possible reason for the multiple function was the authors' inability to easily fine-tune their model. This inability to easily and continuously adapt the performance of the DRL-agent is one of the main novelties of our proposed approach.

To enable a comprehensive and fair comparison, we use the reward functions defined in~\cite{birman2022cost} as our starting point for both of our evaluated datasets. 
Since these functions were not specifically designed for our evaluated datasets, a high performance by \MethodName would further demonstrate its robustness and ability to adapt to changing circumstances.
The five reward functions are presented in Table~\ref{tbls:tbl_experiments}. 
They can generally be divided into two groups---$\{\#1,\#2\}$ and $\{\#3,\#4,\#5\}$---based on their underlying rationale.\\

\begin{comment}
\begin{table}[h]
        \centering
        \small
        %\small
        \setlength\tabcolsep{2pt}
        \begin{tabular}{|c|c|c|c|c|c|}
        \hline
             Experiment & TP & TN & FP & FN\\
            \hline
             1 & $C_{2,j}$ & $C_{2,j}$ & $-C_{2,j}$ & $-C_{2,j}$ \\ 
            \hline
           2 & $C_{2,j}$ & $C_{2,j}$ & $-10C_{2,j}$ & $-10C_{2,j}$ \\
           \hline
            3 & $1$ & $1$ & $-C_{2,j}$ & $-C_{2,j}$ \\ 
            \hline
           4 & $10$ & $10$ & $-C_{2,j}$ & $-C_{2,j}$ \\
           \hline
           5 & $100$ & $100$ & $-C_{2,j}$ & $-C_{2,j}$ \\
           \hline
        \end{tabular}
        \caption{The reward functions used in our experiments are $C_{2,1}=C_{2,2}=C_{2}(t)$. 
        The hyperparameter values for Bahnsen and Wang are $t_2(s) = 0.531$ and $t_2(s) = 0.8925$, respectively.}
    \label{tbls:tbl_experiments}
\end{table}
\end{comment}

\begin{table*}[htp!]
    \fontsize{8}{8}\selectfont
    % \small
    \centering
	\caption{The reward functions used in our experiments are $C_{2,1}=C_{2,2}=C_{2}(t)$. 
        The hyperparameter values for Bahnsen and Wang are $d_2 = 0.531$ and $d_2 = 0.8925$, respectively.
    }
	\label{tbls:tbl_experiments}
    \renewcommand{\arraystretch}{1.1}
	\setlength\tabcolsep{2.7pt}
	\begin{tabular}{ccccc}
		\toprule
    	Experiment & TP & TN & FP & FN\\
		\midrule
        1 & $C_{2,j}$ & $C_{2,j}$ & $-C_{2,j}$ & $-C_{2,j}$ \\ 
            
           2 & $C_{2,j}$ & $C_{2,j}$ & $-10C_{2,j}$ & $-10C_{2,j}$ \\
           
            3 & $1$ & $1$ & $-C_{2,j}$ & $-C_{2,j}$ \\ 
            
           4 & $10$ & $10$ & $-C_{2,j}$ & $-C_{2,j}$ \\
           
           5 & $100$ & $100$ & $-C_{2,j}$ & $-C_{2,j}$ \\
        \bottomrule
	\end{tabular}
\end{table*}

\noindent \textbf{Reward functions \#1 \& \#2.} In these reward functions, both the reward for correct classification and punishment for a mistake are a function of the consumed resources (the two functions differ on the size of the penalty for mis-classification). These reward functions are designed to incentivize the DRL-agent to \textit{place greater emphasis on performance compared to cost}. Because the reward for correct classification is also a function  of consumed resources, the DRL-agent has an incentive to use all available detectors on samples it concludes it has a high probability of classifying correctly. For difficult-to-classify samples, however, the DRL-agent is faced with a balancing-act where it needs to decide how many resources to ``risk'' in an attempt to reach a correct classification.\\

\begin{figure*}[h]
\centering
$\begin{array}{rl}
    \includegraphics[width=0.44 \textwidth]{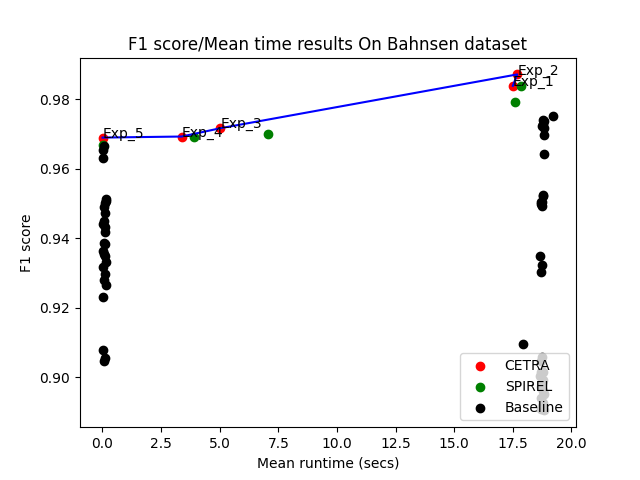}
     & 
    \includegraphics[width=0.44 \textwidth]{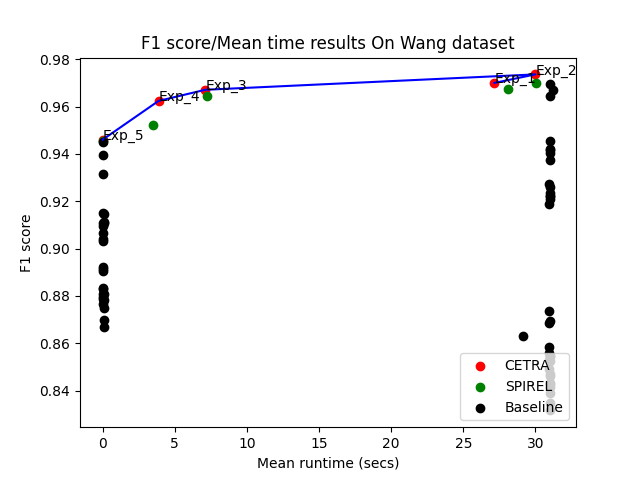} \\
    \includegraphics[width=0.44 \textwidth]{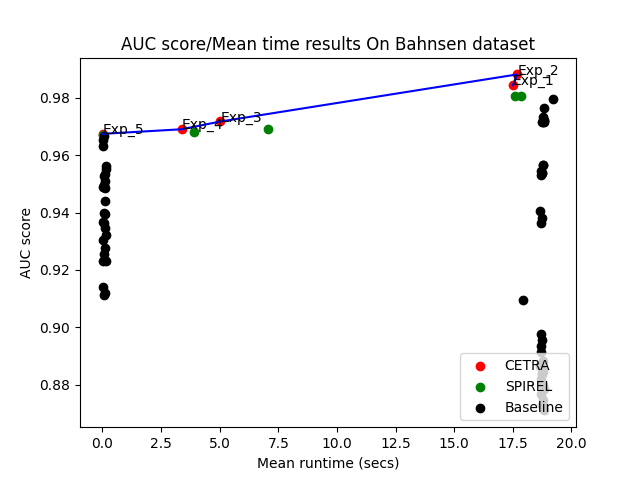} & 
     \includegraphics[width=0.44 \textwidth]{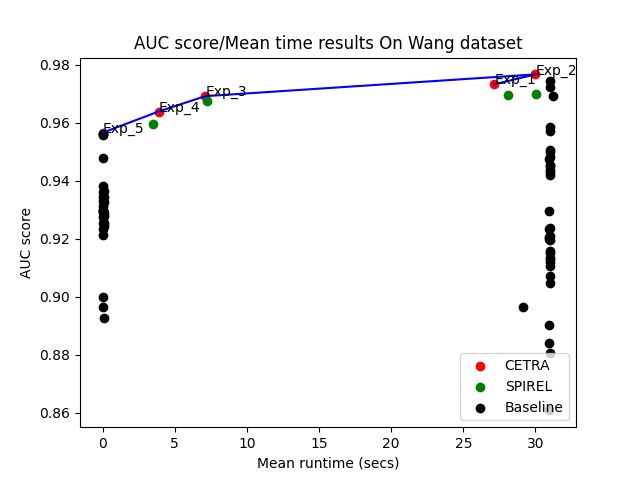}
\end{array}$
\caption{The performance of the baselines on each dataset (F1 and AUC), plotted as a function of the mean time.}
\label{fig:pareto}
\end{figure*}

\noindent \textbf{Reward functions \#3 -- \#5.} In these reward functions, the DRL-agent receives a fixed value for correct classifications, while the cost of an incorrect classification is a function of the computational resources. 
This reward structure incentivizes the DRL-agent to apply as many detectors as it deems necessary to obtain a correct classification, \textit{but not more than that}.
The reason is that any detector that does not contribute new information will not increase the reward in case of a correct classification, but will increase the cost in case of a mistake. 
The DRL-agent must therefore weigh the potential new information a detector can yield against the potential cost of using this detector in case of a mistaken classification.

\section{EVALUATION AND ANALYSIS}

To evaluate \MethodName on our two datasets, we first need to create our algorithm's security policy, i.e., set values for all hyperparameters of our reward function (Equation~\ref{funcs:our_func_total_cur}). 
In Section \ref{subsubsec:PolicyTransfer}, we demonstrate how \textit{we achieve this goal by automatically transferring SPIREL's security policies} using our novel approach, detailed in Section~\ref{subsubsec:KLbasedPolicyTransfer}. 
Next, in Section~\ref{subsubsec:MetricGoalExp}, we demonstrate how \MethodName automatically adapts its security policies to meet organizational goals (e.g., specific recall levels). 

%Our evaluation consists of three parts. 
%In Section~\ref{subsubsec:PolicyTransfer} we evaluate the ability of our density function-based approach to generate a security policy for our evaluated datasets. 
%Next, in Section~\ref{subsubsec:MetricGoalExp} we evaluate the ability of our metric-driven cost function setup to adapt our DRL-agent's behavior to meet the organization's performance goals.
%Finally, in Appendix~\ref{subsec:TransferBasic} we evaluate \MethodName's generalization ability in a ''standard'' transfer learning use-case which we apply after our full approach has been trained. 

\subsection{Security Policy Transfer}
\label{subsubsec:PolicyTransfer}

The parameters of SPIREL's reward function (Equations~\ref{funcs:ref_func_positive} and~\ref{funcs:ref_func}), while effective, were set empirically. 
The parameter $t_1(s)$, which controls the impact of long running times on the cost, was set to the 95th runtime percentile of each detector. 
Our goal is to have \MethodName build on SPIREL's policies, but simply copying the latter's hyperparameters is not likely to yield optimal results. Therefore, we now apply the policy transfer techniques proposed in Section \ref{subsubsec:KLbasedPolicyTransfer} to set the hyperparameter values in Equations \ref{funcs:our_func_main} and \ref{funcs:our_func_positive}.

As described in Section~\ref{subsubsec:KLbasedPolicyTransfer}, our goal is to set a value for the $d_2$ threshold parameter. Our 'starting point' is the function $C_{1}(x)$ (Equation~\ref{funcs:ref_func_positive}), where $d_1 =1$, $s$ is set to the 95th percentile, and $t_1(s) = 34$. Our target function is $C_{2}(x)$ (see Equation~\ref{funcs:our_func_positive}). 
The likelihood ratio $\beta_1$ for reference source $C_{1}(x)$ is $316.5529$. Based on equation~\ref{equations:D_KL_Bi_Bj} and on \ref{inequalities:avg_yi_yj}, the empirically chosen solution of $\beta_2$ is $316.168$ for Bahnsen's dataset, and for Wang's dataset $\beta_2$ is chosen to be $316.232$, where $\varepsilon$ is $10^{-8}$.
Since there is no analytical solution of equation~\ref{integrals:pdf_to_areas_n}, we derived a $d_2$ with a tolerance of $10^{-4}$. 
We found that for Bahnsen's dataset that $d_2$ = $0.531$, and for Wang's dataset that $d_2$ = $0.8925$.

\begin{table*}[htp!]
    \fontsize{6}{8}\selectfont
    % \small
    \centering
	\caption{The top-performing algorithms on the Bahnsen dataset. Note that the complete table is available in Appendix ~\ref{tbls:tbl_detectors_scores_dataset1}.
    }
	\label{tbls:tbl_results_specific_scores_dataset1}
    \renewcommand{\arraystretch}{1.1}
	\setlength\tabcolsep{2.7pt}
	\begin{tabular}{clccccccc}
		\toprule
    	\multirow{2}{*}{} & Combination & Aggregation & AUC & F1 & Time & Precision & Recall & Accuracy\\
		&  &  & (\%) & (\%) & (sec) & (\%) & (\%) & (\%)\\
		\midrule
          & \textbf{\MethodName\_2} & DRL & 98.81 & 98.72 & 17.69 & 98.82 & 98.62 & 98.62 \\ 
          & SPIREL\_2 & DRL & 98.06 & 98.40 & 17.847 & 98.81 & 97.99 & 97.76 \\ 
          & \MethodName\_1 & DRL & 98.43 & 98.39 & 17.5 & 99.47 & 97.32 & 98.49 \\ 
          & SPIREL\_1 & DRL & 98.05 & 97.92 & 17.573 & 98.86 & 96.99 & 97.75 \\ 
         & All Detectors Combined & boosting(ADB) & 97.94 & 97.51 & 19.233 & 96.81 & 98.21 &  97.62 \\ 
         & eXpose, PDRCNN, XGBoost, FFNN & majority & 97.34 & 97.41 & 18.7762 & 97.37 & 97.45 & 97.35 \\ 
         & All Detectors Combined & majority & 97.23 & 97.37 & 18.8163 & 95.91 & 98.87 & 97.27 \\ 
         & eXpose, PDRCNN, CURNN, FFNN & majority & 97.16 & 97.28 & 18.7783 & 96.39 & 98.18 & 97.19 \\ 
         & eXpose, XGBoost, CURNN, FFNN & majority & 97.16 & 97.23 & 18.7457 & 97.21 & 97.25 & 97.16 \\ 
         & PDRCNN, XGBoost, CURNN, FFNN & majority & 97.14 & 97.19 & 18.8135 & 97.66 & 96.72 & 97.13 \\ 
          & \textbf{\MethodName\_3} & DRL & 97.09 & 97.18 & 5.031 & 98.72 & 95.64 & 96.98 \\ 
          & SPIREL\_3 & DRL & 96.91 & 97.00 & 7.069 & 99.03 & 94.97 & 97.04 \\ 
         & All Detectors Combined & stacking(RF) & 97.65 & 96.97 & 18.8169 & 97.80 & 96.15 & 98.31 \\ 
          & \MethodName\_4 & DRL & 96.90 & 96.93 & 3.37 & 98.00 & 95.86 & 97.01 \\ 
          & SPIREL\_4 & DRL & 96.79 & 96.91 & 3.923 & 96.96 & 96.85 & 96.60 \\ 
          & \MethodName\_5 & DRL & 96.74 & 96.90 & 0.002516 & 96.50 & 97.30 & 96.77 \\ 
          & SPIREL\_5 & DRL & 96.71 & 96.68 & 0.00252 & 96.85 & 96.50 & 96.46 \\ 
        \bottomrule
	\end{tabular}
\end{table*}

\begin{table*}[htp!]
    \fontsize{7}{8}\selectfont
    % \small
    \centering
	\caption{The top-performing algorithms on the Wang dataset. Note that the complete table is available in Appendix ~\ref{tbls:tbl_detectors_scores_dataset2}.
    }
	\label{tbls:tbl_results_specific_scores_dataset2}
    \renewcommand{\arraystretch}{1.1}
	\setlength\tabcolsep{2.7pt}
	\begin{tabular}{clccccccc}
		\toprule
    	\multirow{2}{*}{} & Combination & Aggregation & AUC & F1 & Time & Precision & Recall & Accuracy\\
		&  &  & (\%) & (\%) & (sec) & (\%) & (\%) & (\%)\\
		\midrule
          & \textbf{\MethodName\_2} & DRL & 97.66 & 97.36 & 29.993 & 97.98 & 96.74 & 98.27 \\ 
          & SPIREL\_2 & DRL & 96.99 & 97.01 & 30.035 & 98.02 & 96.00 & 97.11  \\
          & \MethodName\_1 & DRL & 97.32 & 97.00 & 27.1499 & 99.60 & 94.40 & 98.23 \\ 
          & All Detectors Combined & stacking(RF) & 97.42 & 96.96 & 31.041 & 98.02 & 95.92 & 98.32 \\ 
          & SPIREL\_1 & DRL & 96.94 & 96.75 & 28.074 & 98.79 & 94.71 & 97.09\\ 
          & All Detectors Combined & boosting(ADB) & 96.93 & 96.71 & 31.247 & 96.26 & 97.16 & 97.11 \\ 
          & \textbf{\MethodName\_3} & DRL & 96.91 & 96.71 & 7.118 & 99.34 & 94.08 & 97.54 \\ 
          & SPIREL\_3 & DRL & 96.73 & 96.46 & 7.216 & 98.92 & 93.99 & 97.02 \\
          & All Detectors Combined & stacking(DT) & 97.22 & 96.44 & 31.0402 & 97.53 & 95.38 & 98.03 \\ 
          & \MethodName\_4 & DRL & 96.38 & 96.24 & 3.886 & 99.50 & 92.99 & 97.55 \\ 
          & SPIREL\_4 & DRL & 95.96 & 95.21 & 3.530 & 99.08 & 91.34 & 96.28  \\
          & \MethodName\_5 & DRL & 95.65 & 94.60 & 0.0075 & 95.81 & 93.39 & 96.21 \\ 
          & SPIREL\_5 & DRL & 95.59 & 94.55 & 0.008 & 95.60 & 93.50 & 96.03 \\ 
        \bottomrule
	\end{tabular}
\end{table*}

We use the approach described above to transfer each of SPIREL's five policies to our two datasets. The results of our evaluation are presented in Tables~\ref{tbls:tbl_results_specific_scores_dataset1} \&~\ref{tbls:tbl_results_specific_scores_dataset2} for the Bahnsen and Wang datasets, respectively. Due to space constraints, we only present the top-performing algorithms for each dataset. The results clearly show the following:

\noindent \textbf{1) \MethodName outperforms all the detector combinations.} In both datasets, \MethodName outperforms the top-performing detector (\MethodName\_2 and \_1) while offering slightly better running times, or achieves slightly lower detection rates (a decrease of 0.5\%-1\% in AUC) while reducing running time by approximately 75\% (\MethodName\_3-5).

\noindent \textbf{2) \MethodName outperforms SPIREL in all policy configurations.} In both datasets, all versions of \MethodName outperform their `origin' policies, achieving both better detection and running times\textit{}. This serves as a clear indication that our policy transfer method can effectively and consistently adapt existing policies to new datasets.\\

\MethodName's clear dominance over SPIREL is important for two reasons. First, it shows that our approach enables `zero-shot' transfer of detection policy, without the need to manually calibrate hyperparameters or run an optimization process. Secondly, the consistent translation demonstrated by our approach---the preservation of the relative ``priorities'' of each origin policy---enable us to create a pool of diverse policies that can be used according to circumstance.

Finally, in Figure~\ref{fig:pareto} we plot all evaluated algorithms (detector combinations as well as all SPIREL and \MethodName configurations) as a function of their performance and mean runtime per sample. 
The charts show that \MethodName creates a \textit{Pareto frontier}, i.e., for every desired level of performance and/or runtime, \MethodName offers the best available solution.

\subsection{Metric-Driven Policy Adaptations} 
\label{subsubsec:MetricGoalExp}

In the previous section we demonstrated our ability to import and adapt successful policies from one dataset to another. We now evaluate our proposed metric-driven cost function approach (see  Section~\ref{subsubsec:MetricCostFunction}), which enables us to refine policies to meet specific organizational goals (e.g., specific TPR rates).

\begin{table*}[htp!]
    \fontsize{8}{8}\selectfont
    % \small
    \centering
	\caption{The results of our proposed approach with and without the metric-driven reward function on the Bahnsen dataset. The former is denoted by ''Metric'', indicating that it's the Metric version of \MethodName. ''Original'' indicates that this configuration only uses the density-based transfer.
    }
	\label{tbls:tbl_tpr_scores_dataset1}
    \renewcommand{\arraystretch}{1.1}
	\setlength\tabcolsep{2.7pt}
	\begin{tabular}{clcccc}
		\toprule
    	\multirow{2}{*}{} & Combination & AUC & F1 & Time & Recall \\
		&   & (\%)  & (\%) & (sec) & (\%)\\
		\midrule
        & \MethodName\_1 Metric &  98.15  & 98.10  & 16.76 & 98.14   \\ 
        & \MethodName\_1 Original &  98.43  & 98.39  & 17.5 & 97.32   \\ 
        & \MethodName\_2 Metric & 98.75  & 98.69  & 17.12  & 98.94  \\ 
        & \MethodName\_2 Original & 98.81  & 98.72  & 17.69 & 98.62  \\ 
        & \MethodName\_3 Metric & 97.04  & 97.09  & 5.012 & 96.21   \\ 
        & \MethodName\_3 Original & 97.09  & 97.18  & 5.031 & 95.64   \\ 
        & \MethodName\_4 Metric & 96.75  & 96.94  & 0.847 & 97.09   \\ 
        & \MethodName\_4 Original & 96.90  & 96.93  & 3.37 & 95.86  \\ 
        & \MethodName\_5 Metric & 96.94  & 96.84  & 0.0026 & 97.39   \\ 
        & \MethodName\_5 Original & 96.74  & 96.90  & 0.0025 & 97.30  \\ 
        \bottomrule
	\end{tabular}
\end{table*}

\begin{table*}[htp!]
    \fontsize{8}{8}\selectfont
    % \small
    \centering
	\caption{The results of our proposed approach with and without the metric-driven reward function on the Wang dataset. 
        former is denoted by "Metric", indicating that it's the Metric version of \MethodName. "Original" indicates that this configuration only uses the density-based transfer.
    }
	\label{tbls:tbl_tpr_scores_dataset2}
    \renewcommand{\arraystretch}{1.1}
	\setlength\tabcolsep{2.7pt}
	\begin{tabular}{clcccc}
		\toprule
    	\multirow{2}{*}{} & Combination & AUC & F1 & Time & Recall \\
		&   & (\%)  & (\%) & (sec) & (\%)\\
		\midrule
        & \MethodName\_1 Metric &  97.73  & 97.18  & 30.057 & 96.41   \\ 
        & \MethodName\_1 Original & 97.32  & 97.00  & 27.1499 & 94.40   \\ 
        & \MethodName\_2 Metric  & 97.61  & 97.29   & 30.19 & 97.02  \\ 
        & \MethodName\_2 Original  & 97.66  & 97.36  & 29.993 & 96.74  \\ 
        & \MethodName\_3 Metric  & 96.84  & 96.52   & 3.155 & 95.54  \\ 
        & \MethodName\_3 Original & 96.91  & 96.71  & 7.118 & 94.08  \\ 
        & \MethodName\_4 Metric &  96.44  & 95.92  & 1.393 & 95.83   \\ 
       &  \MethodName\_4 Original &  96.38  & 96.24  & 3.886 & 92.99  \\ 
        & \MethodName\_5 Metric & 95.50  & 94.56  & 0.002758 & 95.21   \\ 
        & \MethodName\_5 Original &  95.65  & 94.60  & 0.0075 & 93.39  \\ 
        \bottomrule
	\end{tabular}
\end{table*}

For our evaluation, \textit{we chose Recall as our goal metric}. We define the range 95\%-97\% as our ``acceptable'' range, meaning that recall rates above 97\% will provide the agent with a bonus score, while values lower than 95\% will result in a fine. 
We chose these values based on \MethodName's performance on the Wang dataset, where all configurations of our approach except \MethodName\_2 reached recall values below 95\%. 
Because of the very high performance obtained by \MethodName for the Bahnsen dataset, we were not able to define a range of values that is beyond the performance of the existing versions of our approach.

The results of our initial experiments (named 'Original') and the metric-based ones (named 'Metric') for the Bahnsen and Wang datasets (named 'After') are presented in Tables~\ref{tbls:tbl_tpr_scores_dataset1} and~\ref{tbls:tbl_tpr_scores_dataset2}, respectively. 
We focus first on the Wang dataset, where \MethodName performance with respect to the recall metric was not in the specified range in 4 out of 5 configurations. 
The new results clearly show that our metric-driven approach achieved its goals: the recall values of all \MethodName configurations now fall within the specified range. 
Interestingly, the added requirement we placed on our DRL-agent had the effect of providing an additional boost to its performance: both in terms of F1-score and AUC, our approach was able to achieve higher performance compared to the baselines.
Additionally, for two configurations -- \#3 and \#4 -- our metric-driven approach yielded significantly shortened  running times (in the remaining three configurations, the differences were insignificant).

While the performance of \MethodName was already within the predefined recall rang for the Bahnsen dataset, an analysis of the results showed  that our metric-driven approach improved the average running time of our approach in three configurations, slightly extended it in one, and left the remaining configuration unchanged. 
The reduction in running time is particularly notable for configuration \#4, where the average running time dropped from 3.37s to 0.847s, without any impact on the F1 score. 

We attribute this improvement in performance and/or running time to the added incentive our agent now has to exceed the threshold of 97\% recall. Once our approach crosses this line, it receives a ``bonus'' reward to its overall performance. The DRL-agent therefore has the incentive to \textit{slightly exceed the 97\% threshold}, and then direct its attention to improving the overall performance, so it can maximize its bonus. 
A clear example of this is the Bahnsen dataset: the F1 rates of \MethodName\_2 dropped from 98.72\% to 98.69\%, but its recall rates rose from 98.62\% to 98.94\%.

In conclusion, our metric-driven goals are an effective way to create a more nuanced security policy without the need for complex configuration and/or hyper-parameter search. 
Organizations are able to first quantify the actual cost of deploying various detectors, and then add an additional constraints to specify the performance metrics they would like to achieve. 
The DRL-agent then refined its policy to best meet those requirements.

\section{ROBUSTNESS EVALUATION} 
\label{sec:AdvesarialRobustness}

DRL-based solutions have been shown to be vulnerable to adversarial machine learning (AML) attacks~\cite{rosenberg2020adversarial,huang2017adversarial,kos2017delving,lin2017tactics,sun2020stealthy,ilahi2021challenges,tretschk2018sequential}. 
We hypothesize that \MethodName's hierarchical structure, which dynamically selects additional detectors based on the scores of previous ones, would make AML attacks more difficult to execute. 
To test our hypothesis, we evaluate both black-box and white-box attacks. Additionally, because of \MethodName;s focus on efficiency, we evaluate an adversarial attack that seeks to cause larger resource usage.

Note that in all our experiments, \textit{we operate under the strong assumption that the attacker can modify each URL to achieve any desired score from individual detectors}. Under this (highly beneficial to the attacker) assumption, the attack focuses solely on manipulating the DRL agent's `decision process'. For this reason, we expect adversarial attacks to be even less effective in a fully realistic setup.

%we make a strong assumption that, once knowing the required prediction score of a base detector, the attack is able to perturb the url so that the detector will output/provide that score.

\subsection{Evasion -- White-box gradient based attack}
\label{subsec:gradient_based_evasion}

We implemented two variants of an adaptive attack, using two gradient-based white-box attacks: PGD~\cite{madry2017towards} and FSGM~\cite{goodfellow2014explaining}.
The attack is applied to \MethodName as follows:
%Since there is a direct connection between the budget (i.e., epsilon, a value between 0 and 1) of the attack to the crafted adversarial URL, and as the predictions are between 0 and 1, the bigger the budget, the URL will less likely to be phishing. 
%To evaluate our attack, we offer adaptive attack types (gradient and non-gradient attacks) over the $L_{\infty}$ norm. The first is  non-gradient targeted adaptive attack, to evaluate non-gradient ensembles like the baselines. The second is also a white-box targeted adaptive once, but it utilizes the gradients of the DRL to generalise attacks like FSGM and PGD as discussed above.
%In our white-box gradient based adaptive method, we act as follows:
%After training a stochastic policy, in the test stage we always take the most probable action to maximize rewards, which makes the test policy deterministic. 
for each attacked sample, we retrieve the full (pre-attack) trajectory of states $\{s_t\}_{t=1}^T$ chosen by the DRL agent, where $s_t \in S^n$, $S = \{-1\} \cup [0,1]$ and $n$ is the number of base detectors. 
Then, we generate the perturbed trajectory $\{{\hat{s}}_t\}_{t=1}^{\hat{T}}$: in the first timestamp $t=1$, we define ${\hat{s}}_1 = s_1$. 
Then, for each timestamp $t \geq 2$, $\hat{w}_t$ is the result of the targeted PGD/FSGM algorithms on $\hat{s}_{t-1}$, such that for $p = \infty$, $\sum_{t=1}^{\hat{T}-1} \|\hat{s}_t-\hat{w}_t\|_p \leq \varepsilon$. 
%An additional clipping of $\hat{w}_t$ is required, to be a valid state in $S^n$, (see Section~\ref{subsec:baseMethod}). 
Then, ${\hat{s}}_{t}$ is the state reached by following the trained policy $\pi$ on $\hat{w}_t$ and action ${\hat{a}}_{t}$ (i.e., the result of following the perturbed trajectory).

%The main purpose of the attack is to lead to the case that at the last step of the perturbed trajectory $\hat{T}$, ${\hat{a}}_{\hat{T}} \ne a_T$ (i.e., targeting the opposite termination action, as described in section~\ref{subsec:baseMethod}). 
%Our second priority is changing the trajectory $\{s_t\}_{t=1}^T$, which could result in a different path to the opposite termination action. 

\begin{comment}
\begin{equation}
{\hat{s}}_{t+1} = F\left(\pi\left(\hat{w}_{t}\right)\right),\;\; t \geq 1, \;\;{\hat{s}}_1 = s_1
\label{algos:adaptive_gradient_method_drl}
\end{equation}
where $F_{\pi}$ is a function that retrieves the next action from the policy $\pi$ on $\hat{w}_{t}$, and outputs the next state of the trajectory ${\hat{s}}_{t+1}$. 
\end{comment}

Since the number of steps cannot be larger than the number of detectors ${\hat{T}} -1 \leq n$, we choose a sequence $\{\varepsilon_t\}_{t=1}^{n}$, such that at each timestamp $t$, we craft a perturbation $\hat{w}_t$ in $L_{p}$ under the current budget $\varepsilon_t$. 
From that, we can derive the following:
\begin{equation}
\sum_{t=1}^{\hat{T}-1} \|\hat{s}_t-\hat{w}_t\|_p \leq \sum_{t=1}^{n} \|\hat{s}_t-\hat{w}_t\|_p \leq \sum_{t=1}^{n} \varepsilon_t \leq \varepsilon
\label{algos:adaptive_gradient_method_drl}
\end{equation}
where $\sum_{t=1}^{n} \varepsilon_t \leq \varepsilon$ is a necessary condition to meet the requirements of an $L_{p}$ attack under the current budget $\varepsilon$.

We evaluated three attack scenarios: 1) Utilize all the budget at once, defined as: $\varepsilon_t = \varepsilon$ $\textbf{I}_{\{k\}}(t)$, where $k=1,2,...,n$ and $\textbf{I}$ is an indicator function;
2) Utilize the budget uniformly over the trajectory, defined as: $\varepsilon_t = \frac{\varepsilon}{n}$;
3) Utilizes the budget in a geometric decay sequence, defined as: $\varepsilon_t = \varepsilon q^t$, where $q \leq \frac{1}{2}$. 
We use the same two datasets and \MethodName 2 and 3 for our evaluation.
We perform the attacks over the $L_{\infty}$ norm, and use epsilon values (i.e., perturbation size budget) of $0.1, 0.25, 0.5$ for PGD and $0.5$ for FSGM. The number of phishing samples chosen (randomly) for perturbation was $10,000$, with results averaged over three randomly-initiated experiments.

\begin{figure*}[h]
\centering
$\begin{array}{rl}
    \includegraphics[width=0.2 \textwidth]{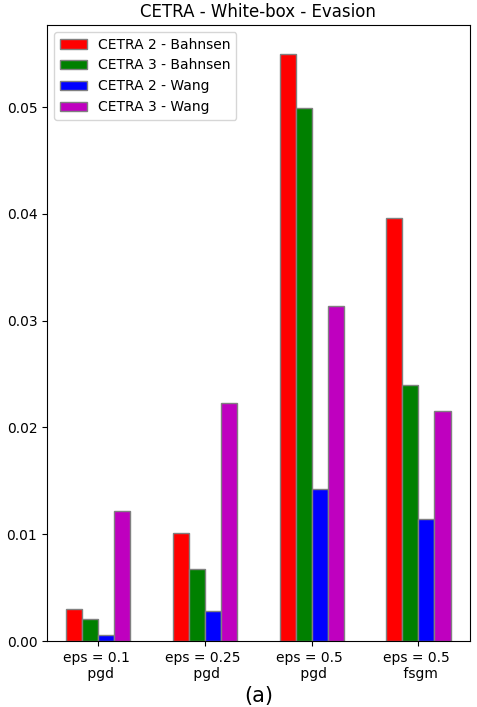}
    \includegraphics[width=0.206 \textwidth]{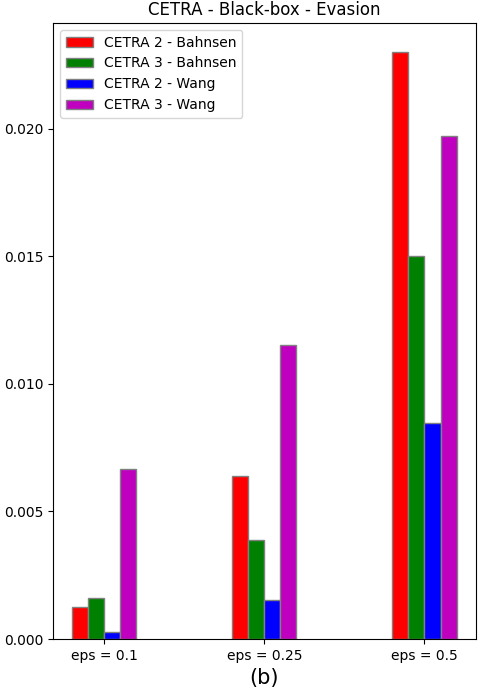}
    \includegraphics[width=0.198 \textwidth]{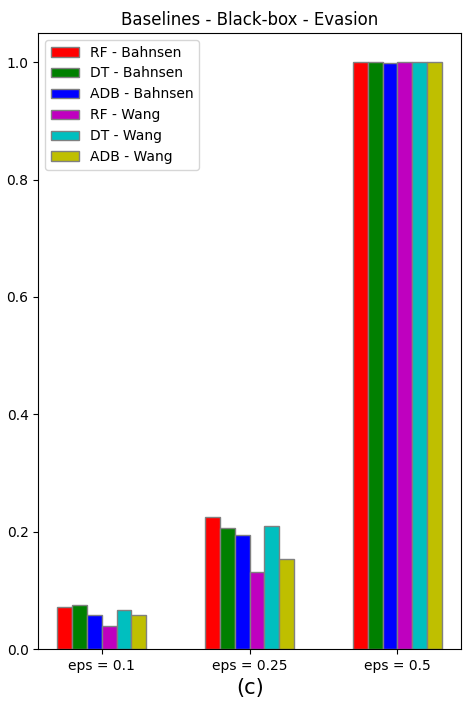}
    \includegraphics[width=0.1995 \textwidth]{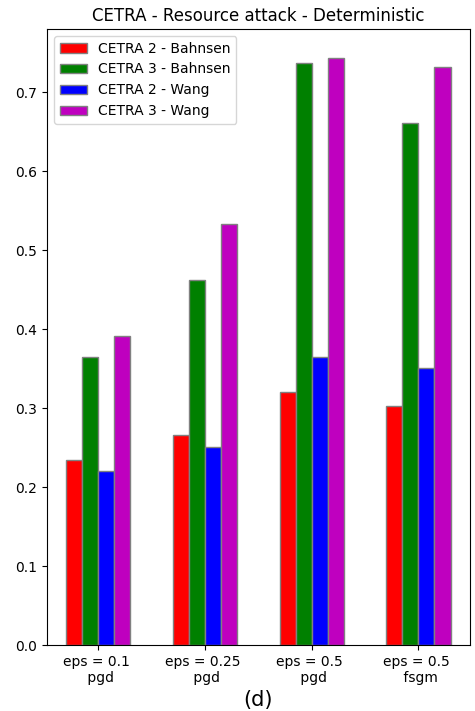}
    \includegraphics[width=0.198 \textwidth]{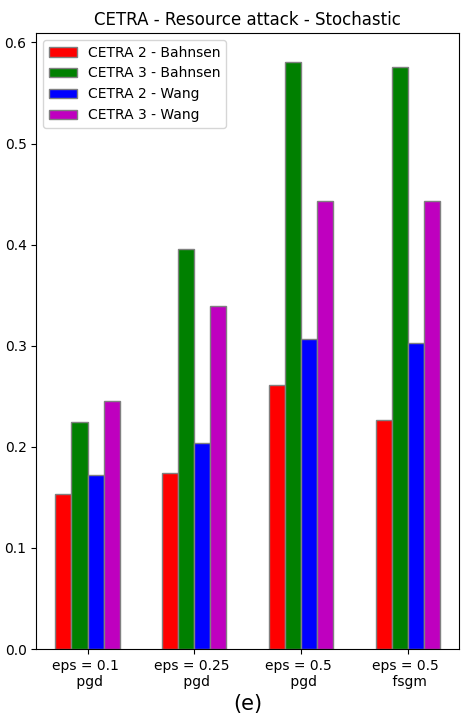}
\end{array}$
\caption{The results of evasion and resource-consumption attacks on \MethodName and the baseline models. We present results for white-box evasion attacks (a), black-box evasion attacks (b-c), and resource consumption attacks (d-e). We present results for \MethodName configurations 2 and 3. please note the different scale of the y-axis of the graphs.}
\label{fig:attack_rates}
\end{figure*}

\begin{comment}
\begin{figure}[t]
    \centering
    \includegraphics[scale=0.52]{figures/misclassify_adversarial_gradient_results.JPG}
    \caption{White-box evasion attack success rate (i.e., classifying malicious URLs as benign). The results are presented for \MethodName\_2 and \MethodName\_3.}
    \label{fig:Phishing_to_Benign_ratio}
\end{figure}
\end{comment}

Our results, presented in Figure~\ref{fig:attack_rates} (a), show that the adversarial attack evade detection by \MethodName in only 5.5\% of the attempts, and under a large epsilon (budget) of 0.5. 
The attack fared worse for lower budgets. 
Additionally, to evade detection, the attack needed to craft a perturbed state in which on average two detector scores are changed, and the magnitude of change (compared with the original state) was 0.07, 0.17 and 0.32 for epsilon 0.1, 0.25 and 0.5, respectively.

These results indicate that \MethodName is highly robust to AML evasion attacks, even without any enhancements. 

\begin{comment}
\begin{table}[h]
        \centering
        \small
        %\setlength\tabcolsep{2pt}
        \begin{tabular}{|c|c|c|c|c|c|c|}
        \hline
        Combination & Attack & Method & Epsilon & Dataset & Rate  \\ \hline
        \MethodName\_2 & White-box &  FSGM & 0.5 & Bahnsen & 3\%  \\ \hline
        \MethodName\_2 & Black-box & Add/Sub & 0.5 & Bahnsen & 3\%  \\ \hline
        \end{tabular} 
        \caption{Attack success rate of white and black-box attacks on \MethodName and the baselines}
    \label{tbls:tbl_attack_evasion}
\end{table}
\end{comment}

%Standard gradient-based adversarial attacks (such as PGD and FSGM) cannot be applied to the baseline ensemble methods used in our research.
%Therefore, in order to evaluate and compare the robustness of these methods to evasion attacks we applied the following attack procedure: 
%In order to evaluate the robustness of baseline ensemble methods  As for the baselines, which are not sequence based algorithms as the DRL, 
%e randomly perturb the input vector $s$ in $L_{p}$ under the budget $\varepsilon$ by adding or subtracting $\varepsilon$ to each feature in the vector for the same budget values: (epsilons), $0.1, 0.25$, and $0.5$.
%The results are presented in Figure~~\ref{fig:Phishing_to_Benign_ratio} (right graph).
%As can be seen, almost 20\% of the attack attempts managed to evade detection by the baseline ensemble methods for epsilon (budget) value of 0.25, and 100\% under a large epsilon (budget) value of 0.5. \gilad{didn't get it}

\subsection{Evasion -- Black-box attack}
\label{subsec:non_gradient_based_evasion}

Gradient-based adversarial attacks (e.g., PGD and FSGM) cannot be applied to our baseline ensemble methods, as they don't use gradients. 
Therefore, we use the following attack procedure to evaluate their and \MethodName's robustness: 
%Therefore, in order to make the comparison of evasion attacks between the robustness of these methods and the DRL even more fair than before, we applied the following attack procedure: 
Instead of perturbing individual states, we randomly perturb the input vector $s$ in $L_{p}$ at each time stamp $t$ using budget $\varepsilon$. 
We do so by adding or subtracting $\varepsilon$ from each feature in the vector.
In essence, we utilize all the budget at once (i.e., $\varepsilon_t = \varepsilon$ $\textbf{I}_{\{k\}}(t)$, where $k=1,2,...,n$ and $\textbf{I}$ is an indicator function). 
For the baselines, which are not sequence-based algorithms, we create $\hat{s}$ by randomly perturbing the input vector $s$ in $L_{p}$  by adding or subtracting $\varepsilon$ to each feature in the vector.

%The attack was evaluated on the two datasets used in this research and on \MethodName's experiments 2 and 3.
%The attacks were performed over the $L_{\infty}$ norm, and the chosen epsilons (i.e., perturbation size budget) were $0.1, 0.25, 0.5$, as we did in the baselines.
%In each experiment, the number of phishing samples chosen (randomly) to perturb into benign is $10,000$ and the results are averaged over three different random sample rounds.

\begin{comment}
\begin{figure}[t]
    \centering
    \includegraphics[scale=0.58]{figures/misclassify_adversarial_non_gradient_results.JPG}
    \caption{Black-box evasion attack success rate (i.e., classifying malicious URLs as benign) by \MethodName\_2 and \MethodName\_3, and compared with the baseline ensembles (please note the different scale of the y-axis of both graphs).}
    \label{fig:Phishing_to_Benign_non_gradient_ratio}
\end{figure}
\end{comment}

We use the same experimental setup as Section~\ref{subsec:gradient_based_evasion}. 
Our results are presented in  Figure~\ref{fig:attack_rates} (b). 
For \MethodName the adversarial attack managed to evade detection in only 2.3\% of the attempts and under a large epsilon value of 0.5---worse results than the white-box attack. 
To evade detection by \MethodName, the attack needed to craft a perturbed state in which on average two detector scores are changed, and the magnitude of change (compared with the original state) is 0.13, 0.19 and 0.41 for epsilon 0.1, 0.25 and 0.5, respectively. 
As for the baselines (Figure~\ref{fig:attack_rates} (c)), almost 20\% of the attack attempts managed to evade detection for $\epsilon$ value of 0.25, and 100\% under $\epsilon=0.5$.

%These results of that black-box attack indicate that \MethodName is even more robust to AML attacks that attempt to evade the detection of phishing URLs, even without any defense mechanism in place.

%Compared to the white-box attack, one intuitive explanation for the difference in the results is that now we do not use different sequences to utilize the budget better, and we do not follow the direction or the opposite direction of the gradient for each different value in the state vector $s$, as in the white-box attack.

%Further evaluation of availability attack based on variations of the DRL attack described above is further analyzed over Appendix~\ref{subsec:Appendix_further_analysis_robustness}.

%, without any additional defense mechanism and with more advanced adaptive gradient based attack, most of the attacks failed. With epsilon 0.5 all the baselines failed to defend on most  of the attacks failed. With epsilon 0.5 all the baselines failed to defense on most of the $10,000$ samples, and  almost only 600 success to overcome \MethodName on its 4 variants on average, both for PGD and FSGM attacks.
%To overcome \MethodName on its 4 variants, the attack need to craft perturb state with almost 2 detectors changes (1.9) on average, and change the magnitude of the detectors which is also activated in the original state by 0.07, 0.17 and 0.32 on epsilon 0.1,0.25 and 0.5, respectively.

\subsection{Resource utilization attack}
\label{subsec:gradient_based_availability}
\MethodName main objective is to reduce the computational cost of detection. 
We now consider adversarial attacks designed to prolong its running time by tricking it into calling additional detectors.
%The attack is successful if it increases the total resource consumption of the DRL agent for that URL, compared to the resource consumption of the original trajectory. 
We modified the attack presented in section~\ref{subsec:gradient_based_evasion} so that from all trajectories that result in the same class, we choose the one where the largest number of detectors were called or with a larger total time than before. 
We use the same setup as in the section~\ref{subsec:gradient_based_evasion}, and evaluate two versions of our approach: a \textit{deterministic case}, where the DRL agent's actions are deterministic, and a \textit{stochastic case} where the probability of choosing an action is proportional to its score.

The results of the attack for the deterministic case are presented in Figure~\ref{fig:attack_rates} (d). For both datasets, the attack succeeded in increasing \MethodName\_2's resource consumption by $33.5\%$ for $\varepsilon = 0.5$, for PGD and FSGM. 
For \MethodName\_3, the attack was more effective, with a $73.5\%$ increase for $\varepsilon = 0.5$. 

The magnitude of sample change needed for a successful attack is 0.08, 0.11 and 0.19 for epsilon 0.1, 0.25 and 0.5, respectively.

The reason for this difference in robustness is \MethodName\_2's  tendency to deploy more detectors due to its focus on detection, whereas \MethodName\_3 gives greater weight to efficiency and tends to call fewer detectors. 

%As the results shown above are high for both datasets, one solution available to reduce the success rate of the attack is to use a stochastic policy also in the evaluation phase~\cite{oikarinen2021robust}. 
%The attack was evaluated almost exactly as mentioned above, but with a tiny change. To address the issue that arises due to the stochasticity, for each URL, we run the attack $10$ times and average the results, thus making them more reliable. 
The results of the attack for the \textit{stochastic case} are presented in Figure~\ref{fig:attack_rates} (e).
Compared to the previous success rate of the deterministic case, the success rate on both datasets for \MethodName\_2 reduced by $7.5\%$ for $\varepsilon = 0.5$, either by using PGD or FSGM. 
As for \MethodName\_3, the attack success rate was reduced even more to $19.5\%$ on average for $\varepsilon = 0.5$. 
The attack needed to craft a perturbed state in which on average two detector scores are changed, and the magnitude of change (compared with the original state) is now 0.12, 0.23 and 0.31 for epsilon 0.1, 0.25 and 0.5, respectively.

The stochastic policy has a minor effect on performance/runtime. 
For the Bahnsen dataset, \MethodName\_2's F1 score was reduced from $98.72\%$ to $98.39\%$, while \MethodName\_3's was reduced from $97.18\%$ to $96.98\%$. 
Similar results were obtained for the Wang dataset. 
The differences in the running time with and without the stochastic policy were negligible---$0.114$ milliseconds on average. 
%To conclude, our results indicate that \MethodName could easily be made more robust---in terms of safeguarding its resource consumption---when using a simple stochastic policy.

\section{CONCLUSIONS}

Utilizing DRL-based solutions in the security domain is challenging both because of the difficulty to calibrate the reward function for new datasets, and because of the need to meet multiple criteria that are possibly not integrated into the reward function. In this study we address both challenges: first, we propose a method of intelligently importing effective security policies from other datasets, thus removing the need for time-consuming optimization. Secondly, we present a method for refining existing policies so that they take into account any additional metrics (and value ranges) desirable to the organization. Evaluation on two real-world datasets shows that our approach is both flexible and effective.

\bibliographystyle{elsarticle-num-names}
\bibliography{references}

\begin{thebibliography}{50}
\expandafter\ifx\csname natexlab\endcsname\relax\def\natexlab#1{#1}\fi
\providecommand{\url}[1]{\texttt{#1}}
\providecommand{\href}[2]{#2}
\providecommand{\path}[1]{#1}
\providecommand{\DOIprefix}{doi:}
\providecommand{\ArXivprefix}{arXiv:}
\providecommand{\URLprefix}{URL: }
\providecommand{\Pubmedprefix}{pmid:}
\providecommand{\doi}[1]{\href{http://dx.doi.org/#1}{\path{#1}}}
\providecommand{\Pubmed}[1]{\href{pmid:#1}{\path{#1}}}
\providecommand{\bibinfo}[2]{#2}
\ifx\xfnm\relax \def\xfnm[#1]{\unskip,\space#1}\fi
%Type = Article
\bibitem[{Ren et~al.(2016)Ren, Zhang, and Suganthan}]{ren2016ensemble}
\bibinfo{author}{Y.~Ren}, \bibinfo{author}{L.~Zhang}, \bibinfo{author}{P.~N.
  Suganthan},
\newblock \bibinfo{title}{Ensemble classification and regression-recent
  developments, applications and future directions},
\newblock \bibinfo{journal}{IEEE Computational intelligence magazine}
  \bibinfo{volume}{11} (\bibinfo{year}{2016}) \bibinfo{pages}{41--53}.
%Type = Inproceedings
\bibitem[{Dvornik et~al.(2019)Dvornik, Schmid, and
  Mairal}]{dvornik2019diversity}
\bibinfo{author}{N.~Dvornik}, \bibinfo{author}{C.~Schmid},
  \bibinfo{author}{J.~Mairal},
\newblock \bibinfo{title}{Diversity with cooperation: Ensemble methods for
  few-shot classification},
\newblock in: \bibinfo{booktitle}{Proceedings of the IEEE/CVF international
  conference on computer vision}, \bibinfo{year}{2019}, pp.
  \bibinfo{pages}{3723--3731}.
%Type = Article
\bibitem[{Belitz and Stackelberg(2021)}]{belitz2021evaluation}
\bibinfo{author}{K.~Belitz}, \bibinfo{author}{P.~Stackelberg},
\newblock \bibinfo{title}{Evaluation of six methods for correcting bias in
  estimates from ensemble tree machine learning regression models},
\newblock \bibinfo{journal}{Environmental Modelling \& Software}
  \bibinfo{volume}{139} (\bibinfo{year}{2021}) \bibinfo{pages}{105006}.
%Type = Article
\bibitem[{Gupta and Gupta(2019)}]{gupta2019dealing}
\bibinfo{author}{S.~Gupta}, \bibinfo{author}{A.~Gupta},
\newblock \bibinfo{title}{Dealing with noise problem in machine learning
  data-sets: A systematic review},
\newblock \bibinfo{journal}{Procedia Computer Science} \bibinfo{volume}{161}
  (\bibinfo{year}{2019}) \bibinfo{pages}{466--474}.
%Type = Article
\bibitem[{Brajard et~al.(2020)Brajard, Carrassi, Bocquet, and
  Bertino}]{brajard2020combining}
\bibinfo{author}{J.~Brajard}, \bibinfo{author}{A.~Carrassi},
  \bibinfo{author}{M.~Bocquet}, \bibinfo{author}{L.~Bertino},
\newblock \bibinfo{title}{Combining data assimilation and machine learning to
  emulate a dynamical model from sparse and noisy observations: A case study
  with the lorenz 96 model},
\newblock \bibinfo{journal}{Journal of Computational Science}
  \bibinfo{volume}{44} (\bibinfo{year}{2020}) \bibinfo{pages}{101171}.
%Type = Article
\bibitem[{Li and Li(2020)}]{li2020adversarial}
\bibinfo{author}{D.~Li}, \bibinfo{author}{Q.~Li},
\newblock \bibinfo{title}{Adversarial deep ensemble: Evasion attacks and
  defenses for malware detection},
\newblock \bibinfo{journal}{IEEE Transactions on Information Forensics and
  Security} \bibinfo{volume}{15} (\bibinfo{year}{2020})
  \bibinfo{pages}{3886--3900}.
%Type = Article
\bibitem[{Yang et~al.(2020)Yang, Zhang, Dong, Inkawhich, Gardner, Touchet,
  Wilkes, Berry, and Li}]{yang2020dverge}
\bibinfo{author}{H.~Yang}, \bibinfo{author}{J.~Zhang},
  \bibinfo{author}{H.~Dong}, \bibinfo{author}{N.~Inkawhich},
  \bibinfo{author}{A.~Gardner}, \bibinfo{author}{A.~Touchet},
  \bibinfo{author}{W.~Wilkes}, \bibinfo{author}{H.~Berry},
  \bibinfo{author}{H.~Li},
\newblock \bibinfo{title}{Dverge: diversifying vulnerabilities for enhanced
  robust generation of ensembles},
\newblock \bibinfo{journal}{Advances in Neural Information Processing Systems}
  \bibinfo{volume}{33} (\bibinfo{year}{2020}) \bibinfo{pages}{5505--5515}.
%Type = Article
\bibitem[{Birman et~al.(2022)Birman, Hindi, Katz, and Shabtai}]{birman2022cost}
\bibinfo{author}{Y.~Birman}, \bibinfo{author}{S.~Hindi},
  \bibinfo{author}{G.~Katz}, \bibinfo{author}{A.~Shabtai},
\newblock \bibinfo{title}{Cost-effective ensemble models selection using deep
  reinforcement learning},
\newblock \bibinfo{journal}{Information Fusion} \bibinfo{volume}{77}
  (\bibinfo{year}{2022}) \bibinfo{pages}{133--148}.
%Type = Article
\bibitem[{Xiang et~al.(2011)Xiang, Hong, Rose, and Cranor}]{xiang2011cantina+}
\bibinfo{author}{G.~Xiang}, \bibinfo{author}{J.~Hong}, \bibinfo{author}{C.~P.
  Rose}, \bibinfo{author}{L.~Cranor},
\newblock \bibinfo{title}{Cantina+ a feature-rich machine learning framework
  for detecting phishing web sites},
\newblock \bibinfo{journal}{ACM Transactions on Information and System Security
  (TISSEC)} \bibinfo{volume}{14} (\bibinfo{year}{2011}) \bibinfo{pages}{1--28}.
%Type = Article
\bibitem[{Horng et~al.(2011)Horng, Fan, Khan, Run, and
  Chen}]{horng2011efficient}
\bibinfo{author}{M.~H.~S. Horng}, \bibinfo{author}{P.~Fan},
  \bibinfo{author}{M.~Khan}, \bibinfo{author}{R.~Run},
  \bibinfo{author}{J.~L.~R. Chen},
\newblock \bibinfo{title}{An efficient phishing webpage detector expertsystems
  with applications},
\newblock \bibinfo{journal}{An International Journal} \bibinfo{volume}{38}
  (\bibinfo{year}{2011}).
%Type = Article
\bibitem[{Zhang et~al.(2014)Zhang, Yan, Jiang, and Kim}]{zhang2014domain}
\bibinfo{author}{D.~Zhang}, \bibinfo{author}{Z.~Yan},
  \bibinfo{author}{H.~Jiang}, \bibinfo{author}{T.~Kim},
\newblock \bibinfo{title}{A domain-feature enhanced classification model for
  the detection of chinese phishing e-business websites},
\newblock \bibinfo{journal}{Information \& Management} \bibinfo{volume}{51}
  (\bibinfo{year}{2014}) \bibinfo{pages}{845--853}.
%Type = Article
\bibitem[{Mohammad et~al.(2014)Mohammad, Thabtah, and
  McCluskey}]{mohammad2014predicting}
\bibinfo{author}{R.~M. Mohammad}, \bibinfo{author}{F.~Thabtah},
  \bibinfo{author}{L.~McCluskey},
\newblock \bibinfo{title}{Predicting phishing websites based on
  self-structuring neural network},
\newblock \bibinfo{journal}{Neural Computing and Applications}
  \bibinfo{volume}{25} (\bibinfo{year}{2014}) \bibinfo{pages}{443--458}.
%Type = Inproceedings
\bibitem[{Verma and Dyer(2015)}]{verma2015character}
\bibinfo{author}{R.~Verma}, \bibinfo{author}{K.~Dyer},
\newblock \bibinfo{title}{On the character of phishing urls: Accurate and
  robust statistical learning classifiers},
\newblock in: \bibinfo{booktitle}{Proceedings of the 5th ACM Conference on Data
  and Application Security and Privacy}, \bibinfo{year}{2015}, pp.
  \bibinfo{pages}{111--122}.
%Type = Article
\bibitem[{Moghimi and Varjani(2016)}]{moghimi2016new}
\bibinfo{author}{M.~Moghimi}, \bibinfo{author}{A.~Y. Varjani},
\newblock \bibinfo{title}{New rule-based phishing detection method},
\newblock \bibinfo{journal}{Expert systems with applications}
  \bibinfo{volume}{53} (\bibinfo{year}{2016}) \bibinfo{pages}{231--242}.
%Type = Article
\bibitem[{Zhang et~al.(2017)Zhang, Jiang, Chen, and Li}]{zhang2017two}
\bibinfo{author}{W.~Zhang}, \bibinfo{author}{Q.~Jiang},
  \bibinfo{author}{L.~Chen}, \bibinfo{author}{C.~Li},
\newblock \bibinfo{title}{Two-stage elm for phishing web pages detection using
  hybrid features},
\newblock \bibinfo{journal}{World Wide Web} \bibinfo{volume}{20}
  (\bibinfo{year}{2017}) \bibinfo{pages}{797--813}.
%Type = Inproceedings
\bibitem[{Bahnsen et~al.(2017)Bahnsen, Bohorquez, Villegas, Vargas, and
  Gonz{\'a}lez}]{bahnsen2017classifying}
\bibinfo{author}{A.~C. Bahnsen}, \bibinfo{author}{E.~C. Bohorquez},
  \bibinfo{author}{S.~Villegas}, \bibinfo{author}{J.~Vargas},
  \bibinfo{author}{F.~A. Gonz{\'a}lez},
\newblock \bibinfo{title}{Classifying phishing urls using recurrent neural
  networks},
\newblock in: \bibinfo{booktitle}{2017 APWG symposium on electronic crime
  research (eCrime)}, \bibinfo{organization}{IEEE}, \bibinfo{year}{2017}, pp.
  \bibinfo{pages}{1--8}.
%Type = Article
\bibitem[{Saxe and Berlin(2017)}]{saxe2017expose}
\bibinfo{author}{J.~Saxe}, \bibinfo{author}{K.~Berlin},
\newblock \bibinfo{title}{expose: A character-level convolutional neural
  network with embeddings for detecting malicious urls, file paths and registry
  keys},
\newblock \bibinfo{journal}{arXiv preprint arXiv:1702.08568}
  (\bibinfo{year}{2017}).
%Type = Article
\bibitem[{Le et~al.(2018)Le, Pham, Sahoo, and Hoi}]{le2018urlnet}
\bibinfo{author}{H.~Le}, \bibinfo{author}{Q.~Pham}, \bibinfo{author}{D.~Sahoo},
  \bibinfo{author}{S.~C. Hoi},
\newblock \bibinfo{title}{Urlnet: Learning a url representation with deep
  learning for malicious url detection},
\newblock \bibinfo{journal}{arXiv preprint arXiv:1802.03162}
  (\bibinfo{year}{2018}).
%Type = Article
\bibitem[{Wang et~al.(2019)Wang, Zhang, Luo, and Zhang}]{wang2019pdrcnn}
\bibinfo{author}{W.~Wang}, \bibinfo{author}{F.~Zhang},
  \bibinfo{author}{X.~Luo}, \bibinfo{author}{S.~Zhang},
\newblock \bibinfo{title}{Pdrcnn: Precise phishing detection with recurrent
  convolutional neural networks},
\newblock \bibinfo{journal}{Security and Communication Networks}
  \bibinfo{volume}{2019} (\bibinfo{year}{2019}).
%Type = Inproceedings
\bibitem[{Kalweit and Boedecker(2017)}]{kalweit2017uncertainty}
\bibinfo{author}{G.~Kalweit}, \bibinfo{author}{J.~Boedecker},
\newblock \bibinfo{title}{Uncertainty-driven imagination for continuous deep
  reinforcement learning},
\newblock in: \bibinfo{booktitle}{Conference on Robot Learning},
  \bibinfo{year}{2017}, pp. \bibinfo{pages}{195--206}.
%Type = Inproceedings
\bibitem[{Van~Hasselt et~al.(2016)Van~Hasselt, Guez, and Silver}]{van2016deep}
\bibinfo{author}{H.~Van~Hasselt}, \bibinfo{author}{A.~Guez},
  \bibinfo{author}{D.~Silver},
\newblock \bibinfo{title}{Deep reinforcement learning with double q-learning},
\newblock in: \bibinfo{booktitle}{Proceedings of the AAAI Conference on
  Artificial Intelligence}, volume~\bibinfo{volume}{30}, \bibinfo{year}{2016}.
%Type = Article
\bibitem[{Mnih et~al.(2015)Mnih, Kavukcuoglu, Silver, Rusu, Veness, Bellemare,
  Graves, Riedmiller, Fidjeland, Ostrovski et~al.}]{mnih2015human}
\bibinfo{author}{V.~Mnih}, \bibinfo{author}{K.~Kavukcuoglu},
  \bibinfo{author}{D.~Silver}, \bibinfo{author}{A.~A. Rusu},
  \bibinfo{author}{J.~Veness}, \bibinfo{author}{M.~G. Bellemare},
  \bibinfo{author}{A.~Graves}, \bibinfo{author}{M.~Riedmiller},
  \bibinfo{author}{A.~K. Fidjeland}, \bibinfo{author}{G.~Ostrovski}, et~al.,
\newblock \bibinfo{title}{Human-level control through deep reinforcement
  learning},
\newblock \bibinfo{journal}{nature} \bibinfo{volume}{518}
  (\bibinfo{year}{2015}) \bibinfo{pages}{529--533}.
%Type = Inproceedings
\bibitem[{Colas et~al.(2018)Colas, Sigaud, and Oudeyer}]{colas2018gep}
\bibinfo{author}{C.~Colas}, \bibinfo{author}{O.~Sigaud}, \bibinfo{author}{P.-Y.
  Oudeyer},
\newblock \bibinfo{title}{Gep-pg: Decoupling exploration and exploitation in
  deep reinforcement learning algorithms},
\newblock in: \bibinfo{booktitle}{International conference on machine
  learning}, \bibinfo{organization}{PMLR}, \bibinfo{year}{2018}, pp.
  \bibinfo{pages}{1039--1048}.
%Type = Inproceedings
\bibitem[{Zhou et~al.(2018)Zhou, Qiao, and Xiang}]{zhou2018deep}
\bibinfo{author}{K.~Zhou}, \bibinfo{author}{Y.~Qiao},
  \bibinfo{author}{T.~Xiang},
\newblock \bibinfo{title}{Deep reinforcement learning for unsupervised video
  summarization with diversity-representativeness reward},
\newblock in: \bibinfo{booktitle}{Proceedings of the AAAI Conference on
  Artificial Intelligence}, volume~\bibinfo{volume}{32}, \bibinfo{year}{2018}.
%Type = Inproceedings
\bibitem[{Feng et~al.(2018)Feng, Huang, Zhao, Yang, and
  Zhu}]{feng2018reinforcement}
\bibinfo{author}{J.~Feng}, \bibinfo{author}{M.~Huang},
  \bibinfo{author}{L.~Zhao}, \bibinfo{author}{Y.~Yang},
  \bibinfo{author}{X.~Zhu},
\newblock \bibinfo{title}{Reinforcement learning for relation classification
  from noisy data},
\newblock in: \bibinfo{booktitle}{Proceedings of the aaai conference on
  artificial intelligence}, volume~\bibinfo{volume}{32}, \bibinfo{year}{2018}.
%Type = Inproceedings
\bibitem[{Blount et~al.(2011)Blount, Tauritz, and Mulder}]{blount2011adaptive}
\bibinfo{author}{J.~J. Blount}, \bibinfo{author}{D.~R. Tauritz},
  \bibinfo{author}{S.~A. Mulder},
\newblock \bibinfo{title}{Adaptive rule-based malware detection employing
  learning classifier systems: a proof of concept},
\newblock in: \bibinfo{booktitle}{2011 IEEE 35th Annual Computer Software and
  Applications Conference Workshops}, \bibinfo{organization}{IEEE},
  \bibinfo{year}{2011}, pp. \bibinfo{pages}{110--115}.
%Type = Article
\bibitem[{Smadi et~al.(2018)Smadi, Aslam, and Zhang}]{smadi2018detection}
\bibinfo{author}{S.~Smadi}, \bibinfo{author}{N.~Aslam},
  \bibinfo{author}{L.~Zhang},
\newblock \bibinfo{title}{Detection of online phishing email using dynamic
  evolving neural network based on reinforcement learning},
\newblock \bibinfo{journal}{Decision Support Systems} \bibinfo{volume}{107}
  (\bibinfo{year}{2018}) \bibinfo{pages}{88--102}.
%Type = Article
\bibitem[{Fu et~al.(2017)Fu, Luo, and Levine}]{fu2017learning}
\bibinfo{author}{J.~Fu}, \bibinfo{author}{K.~Luo}, \bibinfo{author}{S.~Levine},
\newblock \bibinfo{title}{Learning robust rewards with adversarial inverse
  reinforcement learning},
\newblock \bibinfo{journal}{arXiv preprint arXiv:1710.11248}
  (\bibinfo{year}{2017}).
%Type = Article
\bibitem[{Anderson et~al.(2018)Anderson, Kharkar, Filar, Evans, and
  Roth}]{anderson2018learning}
\bibinfo{author}{H.~S. Anderson}, \bibinfo{author}{A.~Kharkar},
  \bibinfo{author}{B.~Filar}, \bibinfo{author}{D.~Evans},
  \bibinfo{author}{P.~Roth},
\newblock \bibinfo{title}{Learning to evade static pe machine learning malware
  models via reinforcement learning},
\newblock \bibinfo{journal}{arXiv preprint arXiv:1801.08917}
  (\bibinfo{year}{2018}).
%Type = Article
\bibitem[{Mo et~al.(2022)Mo, Tang, Li, and Yuan}]{mo2022attacking}
\bibinfo{author}{K.~Mo}, \bibinfo{author}{W.~Tang}, \bibinfo{author}{J.~Li},
  \bibinfo{author}{X.~Yuan},
\newblock \bibinfo{title}{Attacking deep reinforcement learning with decoupled
  adversarial policy},
\newblock \bibinfo{journal}{IEEE Transactions on Dependable and Secure
  Computing}  (\bibinfo{year}{2022}).
%Type = Inproceedings
\bibitem[{Raiber and Kurland(2017)}]{raiber2017kullback}
\bibinfo{author}{F.~Raiber}, \bibinfo{author}{O.~Kurland},
\newblock \bibinfo{title}{Kullback-leibler divergence revisited},
\newblock in: \bibinfo{booktitle}{Proceedings of the ACM SIGIR International
  Conference on Theory of Information Retrieval}, \bibinfo{year}{2017}, pp.
  \bibinfo{pages}{117--124}.
%Type = Article
\bibitem[{Tabibian et~al.(2015)Tabibian, Akbari, and
  Nasersharif}]{tabibian2015speech}
\bibinfo{author}{S.~Tabibian}, \bibinfo{author}{A.~Akbari},
  \bibinfo{author}{B.~Nasersharif},
\newblock \bibinfo{title}{Speech enhancement using a wavelet thresholding
  method based on symmetric kullback--leibler divergence},
\newblock \bibinfo{journal}{Signal Processing} \bibinfo{volume}{106}
  (\bibinfo{year}{2015}) \bibinfo{pages}{184--197}.
%Type = Article
\bibitem[{Jain and Gupta(2019)}]{jain2019machine}
\bibinfo{author}{A.~K. Jain}, \bibinfo{author}{B.~B. Gupta},
\newblock \bibinfo{title}{A machine learning based approach for phishing
  detection using hyperlinks information},
\newblock \bibinfo{journal}{Journal of Ambient Intelligence and Humanized
  Computing} \bibinfo{volume}{10} (\bibinfo{year}{2019})
  \bibinfo{pages}{2015--2028}.
%Type = Article
\bibitem[{Chen et~al.(2015)Chen, He, Benesty, Khotilovich, and
  Tang}]{chen2015xgboost}
\bibinfo{author}{T.~Chen}, \bibinfo{author}{T.~He},
  \bibinfo{author}{M.~Benesty}, \bibinfo{author}{V.~Khotilovich},
  \bibinfo{author}{Y.~Tang},
\newblock \bibinfo{title}{Xgboost: extreme gradient boosting},
\newblock \bibinfo{journal}{R package version 0.4-2}  (\bibinfo{year}{2015})
  \bibinfo{pages}{1--4}.
%Type = Article
\bibitem[{Brockman et~al.(2016)Brockman, Cheung, Pettersson, Schneider,
  Schulman, Tang, and Zaremba}]{brockman2016openai}
\bibinfo{author}{G.~Brockman}, \bibinfo{author}{V.~Cheung},
  \bibinfo{author}{L.~Pettersson}, \bibinfo{author}{J.~Schneider},
  \bibinfo{author}{J.~Schulman}, \bibinfo{author}{J.~Tang},
  \bibinfo{author}{W.~Zaremba},
\newblock \bibinfo{title}{Openai gym},
\newblock \bibinfo{journal}{arXiv preprint arXiv:1606.01540}
  (\bibinfo{year}{2016}).
%Type = Article
\bibitem[{Lin(1992)}]{lin1992self}
\bibinfo{author}{L.-J. Lin},
\newblock \bibinfo{title}{Self-improving reactive agents based on reinforcement
  learning, planning and teaching},
\newblock \bibinfo{journal}{Machine learning} \bibinfo{volume}{8}
  (\bibinfo{year}{1992}) \bibinfo{pages}{293--321}.
%Type = Incollection
\bibitem[{Nandy and Biswas(2018)}]{nandy2018reinforcement}
\bibinfo{author}{A.~Nandy}, \bibinfo{author}{M.~Biswas},
\newblock \bibinfo{title}{Reinforcement learning with keras, tensorflow, and
  chainerrl},
\newblock in: \bibinfo{booktitle}{Reinforcement Learning},
  \bibinfo{publisher}{Springer}, \bibinfo{year}{2018}, pp.
  \bibinfo{pages}{129--153}.
%Type = Article
\bibitem[{Tieleman and Hinton(2012)}]{tieleman2012lecture}
\bibinfo{author}{T.~Tieleman}, \bibinfo{author}{G.~Hinton},
\newblock \bibinfo{title}{Lecture 6.5-rmsprop: Divide the gradient by a running
  average of its recent magnitude},
\newblock \bibinfo{journal}{COURSERA: Neural networks for machine learning}
  \bibinfo{volume}{4} (\bibinfo{year}{2012}) \bibinfo{pages}{26--31}.
%Type = Inproceedings
\bibitem[{Zhang et~al.(2015)Zhang, Wang, and Zhao}]{zhang2015estimating}
\bibinfo{author}{D.~Zhang}, \bibinfo{author}{J.~Wang},
  \bibinfo{author}{X.~Zhao},
\newblock \bibinfo{title}{Estimating the uncertainty of average f1 scores},
\newblock in: \bibinfo{booktitle}{Proceedings of the 2015 International
  conference on the theory of information retrieval}, \bibinfo{year}{2015}, pp.
  \bibinfo{pages}{317--320}.
%Type = Article
\bibitem[{Rosenberg et~al.(2020)Rosenberg, Shabtai, Elovici, and
  Rokach}]{rosenberg2020adversarial}
\bibinfo{author}{I.~Rosenberg}, \bibinfo{author}{A.~Shabtai},
  \bibinfo{author}{Y.~Elovici}, \bibinfo{author}{L.~Rokach},
\newblock \bibinfo{title}{Adversarial learning in the cyber security domain}
  (\bibinfo{year}{2020}).
%Type = Article
\bibitem[{Huang et~al.(2017)Huang, Papernot, Goodfellow, Duan, and
  Abbeel}]{huang2017adversarial}
\bibinfo{author}{S.~Huang}, \bibinfo{author}{N.~Papernot},
  \bibinfo{author}{I.~Goodfellow}, \bibinfo{author}{Y.~Duan},
  \bibinfo{author}{P.~Abbeel},
\newblock \bibinfo{title}{Adversarial attacks on neural network policies},
\newblock \bibinfo{journal}{arXiv preprint arXiv:1702.02284}
  (\bibinfo{year}{2017}).
%Type = Article
\bibitem[{Kos and Song(2017)}]{kos2017delving}
\bibinfo{author}{J.~Kos}, \bibinfo{author}{D.~Song},
\newblock \bibinfo{title}{Delving into adversarial attacks on deep policies},
\newblock \bibinfo{journal}{arXiv preprint arXiv:1705.06452}
  (\bibinfo{year}{2017}).
%Type = Article
\bibitem[{Lin et~al.(2017)Lin, Hong, Liao, Shih, Liu, and Sun}]{lin2017tactics}
\bibinfo{author}{Y.-C. Lin}, \bibinfo{author}{Z.-W. Hong},
  \bibinfo{author}{Y.-H. Liao}, \bibinfo{author}{M.-L. Shih},
  \bibinfo{author}{M.-Y. Liu}, \bibinfo{author}{M.~Sun},
\newblock \bibinfo{title}{Tactics of adversarial attack on deep reinforcement
  learning agents},
\newblock \bibinfo{journal}{arXiv preprint arXiv:1703.06748}
  (\bibinfo{year}{2017}).
%Type = Inproceedings
\bibitem[{Sun et~al.(2020)Sun, Zhang, Xie, Ma, Zheng, Chen, and
  Liu}]{sun2020stealthy}
\bibinfo{author}{J.~Sun}, \bibinfo{author}{T.~Zhang}, \bibinfo{author}{X.~Xie},
  \bibinfo{author}{L.~Ma}, \bibinfo{author}{Y.~Zheng},
  \bibinfo{author}{K.~Chen}, \bibinfo{author}{Y.~Liu},
\newblock \bibinfo{title}{Stealthy and efficient adversarial attacks against
  deep reinforcement learning},
\newblock in: \bibinfo{booktitle}{Proceedings of the AAAI Conference on
  Artificial Intelligence}, volume~\bibinfo{volume}{34}, \bibinfo{year}{2020},
  pp. \bibinfo{pages}{5883--5891}.
%Type = Article
\bibitem[{Ilahi et~al.(2021)Ilahi, Usama, Qadir, Janjua, Al-Fuqaha, Huang, and
  Niyato}]{ilahi2021challenges}
\bibinfo{author}{I.~Ilahi}, \bibinfo{author}{M.~Usama},
  \bibinfo{author}{J.~Qadir}, \bibinfo{author}{M.~U. Janjua},
  \bibinfo{author}{A.~Al-Fuqaha}, \bibinfo{author}{D.~T. Huang},
  \bibinfo{author}{D.~Niyato},
\newblock \bibinfo{title}{Challenges and countermeasures for adversarial
  attacks on deep reinforcement learning},
\newblock \bibinfo{journal}{IEEE Transactions on Artificial Intelligence}
  (\bibinfo{year}{2021}).
%Type = Article
\bibitem[{Tretschk et~al.(2018)Tretschk, Oh, and
  Fritz}]{tretschk2018sequential}
\bibinfo{author}{E.~Tretschk}, \bibinfo{author}{S.~J. Oh},
  \bibinfo{author}{M.~Fritz},
\newblock \bibinfo{title}{Sequential attacks on agents for long-term
  adversarial goals},
\newblock \bibinfo{journal}{arXiv preprint arXiv:1805.12487}
  (\bibinfo{year}{2018}).
%Type = Article
\bibitem[{Madry et~al.(2017)Madry, Makelov, Schmidt, Tsipras, and
  Vladu}]{madry2017towards}
\bibinfo{author}{A.~Madry}, \bibinfo{author}{A.~Makelov},
  \bibinfo{author}{L.~Schmidt}, \bibinfo{author}{D.~Tsipras},
  \bibinfo{author}{A.~Vladu},
\newblock \bibinfo{title}{Towards deep learning models resistant to adversarial
  attacks},
\newblock \bibinfo{journal}{arXiv preprint arXiv:1706.06083}
  (\bibinfo{year}{2017}).
%Type = Article
\bibitem[{Goodfellow et~al.(2014)Goodfellow, Shlens, and
  Szegedy}]{goodfellow2014explaining}
\bibinfo{author}{I.~J. Goodfellow}, \bibinfo{author}{J.~Shlens},
  \bibinfo{author}{C.~Szegedy},
\newblock \bibinfo{title}{Explaining and harnessing adversarial examples},
\newblock \bibinfo{journal}{arXiv preprint arXiv:1412.6572}
  (\bibinfo{year}{2014}).
%Type = Article
\bibitem[{Markechov{\'a}(2017)}]{markechova2017kullback}
\bibinfo{author}{D.~Markechov{\'a}},
\newblock \bibinfo{title}{Kullback-leibler divergence and mutual information of
  experiments in the fuzzy case},
\newblock \bibinfo{journal}{Axioms} \bibinfo{volume}{6} (\bibinfo{year}{2017})
  \bibinfo{pages}{5}.
%Type = Book
\bibitem[{Baccelli and Br{\'e}maud(2012)}]{baccelli2012palm}
\bibinfo{author}{F.~Baccelli}, \bibinfo{author}{P.~Br{\'e}maud},
  \bibinfo{title}{Palm probabilities and stationary queues},
  volume~\bibinfo{volume}{41}, \bibinfo{publisher}{Springer Science \& Business
  Media}, \bibinfo{year}{2012}.

\end{thebibliography}

\clearpage
%\newpage
\appendix
%\section{APPENDIX}

\section{Detector Performance Analysis}
\label{subsec:detectorPerformance}
For an ensemble to perform well, the detectors need to have \textit{meaningful differences in performance}, meaning that one detector is capable of correctly classifying samples another detector could not.
Moreover, it is important to ensure that no detector ``dominates'' another, i.e., achieves an equal or more accurate classification for every sample. 
Our analysis in this section demonstrates that these two conditions are met.
Throughout this section, we use standard practices and evaluation metrics. 
Given that all our detectors output classifications in the range of $[0,1]$, we use the following setting to determine the classification of each sample:

\begin{equation}
 p_{i} =
    \begin{cases}
      benign & ,\text{$ prediction \leq 0.5 $}\\
      phishing & ,\text{$ otherwise $}
    \end{cases}
    \label{formulas:thresholds}
\end{equation}

\begin{table*}[htp!]
    \fontsize{8}{8}\selectfont
    % \small
    \centering
	\caption{The performance of the individual detectors on the Bahnsen dataset.
    }
	\label{tbls:tbl_basic_detectors_scores_dataset1}
    \renewcommand{\arraystretch}{1.1}
	\setlength\tabcolsep{2.7pt}
	\begin{tabular}{clcccccc}
		\toprule
    	\multirow{2}{*}{} & Detector & AUC & F1 & Time & Precision & Recall & Accuracy \\
		&   & (\%)  & (\%) & (sec) & (\%) & (\%) & (\%)\\
		\midrule
         & CURNN & 96.42 & 96.31 & 0.0400 & 95.71 & 96.91 & 96.40 \\ 
         & eXpose & 96.58 & 96.53 & 0.0025 & 94.31 & 98.75 & 96.51 \\ 
         & XGBoost & 92.58 & 92.30 & 0.0379 & 95.51 & 89.09 & 92.69 \\ 
         & PDRCNN & 96.76 & 96.66 & 0.0609 & 96.16 & 97.17 & 96.75 \\ 
         & FFNN & 90.96 & 90.96 & 17.9453 & 87.97 & 93.95 & 90.86 \\ 
        \bottomrule
	\end{tabular}
\end{table*}

\begin{table*}[htp!]
    \fontsize{8}{8}\selectfont
    % \small
    \centering
	\caption{The performance of the individual detectors on the Wang dataset
    }
	\label{tbls:tbl_basic_detectors_scores_dataset2}
    \renewcommand{\arraystretch}{1.1}
	\setlength\tabcolsep{2.7pt}
	\begin{tabular}{clcccccc}
		\toprule
    	\multirow{2}{*}{} & Detector & AUC & F1 & Time & Precision & Recall & Accuracy \\
		&   & (\%)  & (\%) & (sec) & (\%) & (\%) & (\%)\\
		\midrule
        & CURNN &  95.62 & 94.52 & 0.01831 & 95.40 & 93.63 & 96.23 \\ 
        & eXpose & 95.56 & 93.94 & 0.0024 & 92.94 & 94.95 & 95.75 \\ 
        & XGBoost &  92.54 & 90.65 & 0.0377 & 92.23 & 89.06 & 93.61 \\ 
        & PDRCNN & 94.78 & 93.17 & 0.0390 & 93.14 & 93.20 & 95.26 \\ 
        & FFNN &  89.65 & 86.29 & 29.1686 & 85.24 & 87.34 & 90.36 \\ 
        \bottomrule
	\end{tabular}
\end{table*}

The performance of our detectors for the two datasets described in Section~\ref{subsec:datasets} is presented in Tables~\ref{tbls:tbl_basic_detectors_scores_dataset1} \&~\ref{tbls:tbl_basic_detectors_scores_dataset2}. 
It is clear that all detectors have high yet diverse rates of precision and recall, a fact that satisfies our first requirement. 
Moreover, it is clear that there is large variance in the detectors' average running time per sample, a fact that makes the selection of a subset of them potentially very cost-effective.

Next, we analyze the performance of the detectors to determine whether the performance of any of them is dominated by another. 
Our analysis is performed as follows: for each detector, we only consider the samples that were \textit{classified incorrectly}. 
Next, for these samples, we analyze the performance of all other detectors to determine for each of them whether it is able to correctly classify at least some of them. 
If that is the case, we can conclude that the latter detectors are not being dominated by the former.
The results of this analysis are presented in Tables~\ref{tbls:tbl_domination_dataset1} \&~\ref{tbls:tbl_domination_dataset2}. 
It is clear that no detector dominates another, and that the ability of one detector to compensate for the failings of another is often high.
It should also be noted that FFNN, the detector whose performance was a little lower than those of the other detectors (see Tables~\ref{tbls:tbl_basic_detectors_scores_dataset1} \&~\ref{tbls:tbl_basic_detectors_scores_dataset2}), has the best average performance in this analysis.

To conclude, in this section we were able to show that not only do our detectors have high performance rates in their own right, combining them in an ensemble has the potential to yield non-trivial improvement in performance. 
In the following section we describe how \MethodName can be used to improve upon the performance of the standard ensemble.

\begin{table*}[htp!]
    \fontsize{8}{8}\selectfont
    % \small
    \centering
	\caption{Percentage of URLs mis-classified by one detector (row) and correctly classified by another (column) -- Bahnsen dataset.
    }
	\label{tbls:tbl_domination_dataset1}
    \renewcommand{\arraystretch}{1.1}
	\setlength\tabcolsep{2.7pt}
	\begin{tabular}{clccccc}
		\toprule
    	\multirow{2}{*}{} & Method & CURNN & eXpose & XGBoost & PDRCNN & FFNN  \\
	& & (\%)  & (\%) & (\%) & (\%) & (\%)\\
		\midrule
       & CURNN & \textbf{-} & 41.94 & 62.6  & 48.41 & 69.81   \\
             
            & eXpose & 45.36  & \textbf{-} & 71.34  & 45.76 & 66.45   \\
            
          &   XGBoost & 81.38  & 84.83  & \textbf{-} & 84.86 & 81.03  \\
            
          &   PDRCNN & 43.36  & 36.71  & 66.61 &  \textbf{-} & 71.11   \\
            
          &   FFNN & 87.35  & 85.06  & 84.04 & 88.97 & \textbf{-}  \\
        \bottomrule
	\end{tabular}
\end{table*}
\begin{table*}[htp!]
    \fontsize{8}{8}\selectfont
    % \small
    \centering
	\caption{Percentage of URLs mis-classified by one detector (row) and correctly classified by another (column) -- Wang dataset.
    }
	\label{tbls:tbl_domination_dataset2}
    \renewcommand{\arraystretch}{1.1}
	\setlength\tabcolsep{2.7pt}
	\begin{tabular}{clccccc}
		\toprule
    	\multirow{2}{*}{} & Method & CURNN & eXpose & XGBoost & PDRCNN & FFNN  \\
	& & (\%)  & (\%) & (\%) & (\%) & (\%)\\
		\midrule
       & CURNN & \textbf{-} & 41.56 & 45.46  & 36.7 & 84.5   \\
                         
        & eXpose & 49.93  & \textbf{-} & 57.84  & 45.53 & 85.57   \\
            
          &   XGBoost & 65.20  & 68.60  & \textbf{-} & 63.27 & 86.37  \\
            
            & PDRCNN & 48.76  & 48.53  & 53.39 &  \textbf{-} & 85.39   \\
            
           &  FFNN & 93.09  & 92.59  & 90.48 & 91.96 & \textbf{-}  \\ 
        \bottomrule
	\end{tabular}
\end{table*}

\section{Dataset transferability experiment} 
\label{subsec:TransferBasic}

The goal of this set of experiments is to evaluate the robustness and transfer-ability of the policies generated by our proposed approach. 
To this end, we train our full model (using the density-based transfer learning + metric-driven cost function) on the training set of one of our datasets, then apply it to the test set of another. 

The results of our evaluation are presented in Table~\ref{tbls:tbl_results_train_test_mixes}.
The results clearly show that our generated policies perform very well when applied to different datasets with similar characteristics. 
It should also be pointed out that our various reward function configurations maintain the same relative ranking in the performance/runtime trade-offs, further demonstrating the robustness and stability of our approach.

\begin{table*}[htp!]
    \fontsize{8}{8}\selectfont
    % \small
    \centering
	\caption{\MethodName's performance in a transfer-learning setting. We train our approach on one dataset, then apply it to the other.
    }
	\label{tbls:tbl_results_train_test_mixes}
    \renewcommand{\arraystretch}{1.1}
	\setlength\tabcolsep{2.7pt}
	\begin{tabular}{ccclcccccc}
		\toprule
    	\multirow{2}{*}{} & Train set & Test set & Combination & AUC & F1 & Time & Precision & Recall & Accuracy \\
	&   & 	&   & (\%)  & (\%) & (sec) & (\%) & (\%) & (\%)\\
		\midrule
         & Wang & Bahnsen & \textbf{\MethodName\_2} & 99.08 & 99.04 & 18.57 & 99.27 & 98.82 & 99.10 \\ 
        & Wang & Bahnsen & \MethodName\_1 & 98.76 & 98.71 & 17.32 & 98.87 & 98.55 & 98.77 \\ 
        & Wang & Bahnsen & \textbf{\MethodName\_3} & 98.51 & 98.46 & 5.141 & 99.03 & 97.89 & 98.53 \\ 
        & Wang & Bahnsen & \MethodName\_4 & 98.00 & 97.71 & 2.74 & 96.59 & 98.86 & 97.91 \\ 
        & Wang & Bahnsen & \MethodName\_5 & 96.90 & 96.80 & 0.0057 & 95.55 & 98.09 & 96.86 \\ 
        & Bahnsen & Wang & \textbf{\MethodName\_2} & 96.88 & 96.60 & 27.641 & 99.20 & 94.14 & 97.82 \\ 
        & Bahnsen & Wang & \MethodName\_1 & 96.72 & 95.61 & 27.441 & 95.15 & 96.07 & 96.92 \\ 
        & Bahnsen & Wang & \textbf{\MethodName\_3} & 95.30 & 95.03 & 7.666 & 99.82 & 90.67 & 96.95 \\ 
        & Bahnsen & Wang & \MethodName\_4 & 94.89 & 94.18 & 2.97 & 97.24 & 91.30 & 95.84 \\ 
        & Bahnsen & Wang & \MethodName\_5 & 94.40 & 93.89 & 0.00238 & 98.90 & 89.37 & 95.95 \\  
        \bottomrule
	\end{tabular}
\end{table*}

\section{Full Results}
\label{subsec:Appendix_full_results_tables}
The full results of our evaluation are presented in Tables~\ref{tbls:tbl_detectors_scores_dataset1} \&~\ref{tbls:tbl_detectors_scores_dataset2} for the Bahnsen and Wang datasets, respectively.

\clearpage
\thispagestyle{empty}

\begin{table*}[htp!]
    \fontsize{6}{8}\selectfont
    % \small
    \centering
	\caption{Full description of methods' results on Bahnsen's dataset (ordered based on the F1 score).
    }
	\label{tbls:tbl_detectors_scores_dataset1}
    \renewcommand{\arraystretch}{1.1}
	\setlength\tabcolsep{2.7pt}
	\begin{tabular}{clccccccc}
		\toprule
    	\multirow{2}{*}{} & Combination & Aggregation & AUC & F1 & Time & Precision & Recall & Accuracy\\
		&  & & (\%) & (\%) & (sec) & (\%) & (\%) & (\%)\\
		\midrule
          & \textbf{\MethodName\_2} & DRL & 98.81 & 98.72 & 17.69 & 98.82 & 98.62 & 98.62 \\ 
          & SPIREL\_2 & DRL & 98.06 & 98.40 & 17.847 & 98.81 & 97.99 & 97.76 \\ 
          & \MethodName\_1 & DRL & 98.43 & 98.39 & 17.5 & 99.47 & 97.32 & 98.49 \\ 
          & SPIREL\_1 & DRL & 98.05 & 97.92 & 17.573 & 98.86 & 96.99 & 97.75 \\ 
            & All Detectors Combined & boosting(ADB) & 97.94 & 97.51 & 19.233 & 96.81 & 98.21 &  97.62 \\ 
            & eXpose, PDRCNN, XGBoost, FFNN & majority & 97.34 & 97.41 & 18.7762 & 97.37 & 97.45 & 97.35 \\ 
            & All Detectors Combined & majority & 97.23 & 97.37 & 18.8163 & 95.91 & 98.87 & 97.27 \\ 
            & eXpose, PDRCNN, CURNN, FFNN & majority & 97.16 & 97.28 & 18.7783 & 96.39 & 98.18 & 97.19 \\ 
            & eXpose, XGBoost, CURNN, FFNN & majority & 97.16 & 97.23 & 18.7457 & 97.21 & 97.25 & 97.16 \\ 
            & PDRCNN, XGBoost, CURNN, FFNN & majority & 97.14 & 97.19 & 18.8135 & 97.66 & 96.72 & 97.13 \\ 
          & \textbf{\MethodName\_3} & DRL & 97.09 & 97.18 & 5.031 & 98.72 & 95.64 & 96.98 \\ 
          & SPIREL\_3 & DRL & 96.91 & 97.00 & 7.069 & 99.03 & 94.97 & 97.04 \\ 
            & All Detectors Combined & stacking(RF) & 97.65 & 96.97 & 18.8169 & 97.80 & 96.15 & 98.31 \\ 
          & \MethodName\_4 & DRL & 96.90 & 96.93 & 3.37 & 98.00 & 95.86 & 97.01 \\ 
          & SPIREL\_4 & DRL & 96.79 & 96.91 & 3.923 & 96.96 & 96.85 & 96.60 \\ 
          & \MethodName\_5 & DRL & 96.74 & 96.90 & 0.002516 & 96.50 & 97.30 & 96.77 \\ 
          & SPIREL\_5 & DRL & 96.71 & 96.68 & 0.00252 & 96.85 & 96.50 & 96.46 \\ 
        %%TOREMOVE-PDRCNN & none & 96.76 & 96.66 & 0.0609 & 96.16 & 97.17 & 96.75 \\ 
        %%TOREMOVE-eXpose & none & 96.58 & 96.53 & 0.0025 & 94.31 & 98.75 & 96.51 \\ 
            & All Detectors Combined & stacking(DT) & 97.17 & 96.42 & 18.8165 & 97.63 & 95.24 & 98.02 \\ 
        %%TOREMOVE-CURNN & none & 96.42 & 96.31 & 0.0400 & 95.71 & 96.91 & 96.40 \\ 
            & PDRCNN, CURNN, FFNN & majority & 95.66 & 95.24 & 18.7730 & 94.97 & 95.51 & 95.67 \\ 
            & PDRCNN, XGBoost, FFNN & majority & 95.67 & 95.21 & 18.7745 & 95.44 & 94.98 & 95.66 \\ 
            & eXpose, PDRCNN, XGBoost, CURNN & majority & 95.61 & 95.13 & 0.1510 & 95.77 & 94.48 & 95.59 \\ 
            & PDRCNN, XGBoost, CURNN & majority & 95.52 & 95.06 & 0.1485 & 95.34 & 94.77 & 95.51 \\ 
            & eXpose, XGBoost, FFNN & majority & 95.44 & 95.05 & 18.7070 & 94.58 & 95.51 & 95.45 \\ 
            & eXpose, PDRCNN, FFNN & majority & 95.37 & 95.03 & 18.7388 & 94.06 & 95.99 & 95.39 \\ 
            & eXpose, PDRCNN, XGBoost & majority & 95.34 & 95.02 & 0.1121 & 94.53 & 95.51 & 95.42 \\ 
            & eXpose, CURNN, FFNN & majority & 95.32 & 94.98 & 18.7080 & 93.94 & 96.01 & 95.34 \\ 
            & XGBoost, CURNN, FFNN & majority & 95.38 & 94.93 & 18.7443 & 95.13 & 94.73 & 95.37 \\ 
            & eXpose, XGBoost, CURNN & majority & 95.28 & 94.89 & 0.0802 & 94.49 & 95.29 & 95.29 \\ 
            & eXpose, PDRCNN, CURNN & majority & 95.10 & 94.74 & 0.1130 & 94.00 & 95.48 & 95.12 \\ 
            & eXpose, PDRCNN & majority & 94.97 & 94.50 & 0.0732 & 95.20 & 93.79 & 94.95 \\ 
            & eXpose, CURNN & majority & 94.88 & 94.41 & 0.0424 & 95.13 & 93.68 & 94.86 \\ 
            & PDRCNN, CURNN & majority & 94.85 & 94.32 & 0.1107 & 96.03 & 92.60 & 94.80 \\ 
            & PDRCNN, CURNN & or & 94.40 & 94.18 & 0.1104 & 92.52 & 95.83 & 94.44 \\ 
            & eXpose, PDRCNN & or & 93.97 & 93.87 & 0.0731 & 91.50 & 96.23 & 94.03 \\ 
            & PDRCNN, XGBoost & or & 93.94 & 93.83 & 0.1085 & 91.57 & 96.08 & 94.00 \\ 
            & eXpose, CURNN & or & 93.68 & 93.64 & 0.0423 & 91.00 & 96.27 & 93.75 \\ 
            & XGBoost, CURNN & or & 93.64 & 93.55 & 0.0777 & 91.26 & 95.83 & 93.70 \\ 
            & eXpose, PDRCNN, CURNN & or & 93.47 & 93.50 & 0.1129 & 90.56 & 96.43 & 93.55 \\ 
            & eXpose, FFNN & majority & 94.04 & 93.48 & 18.6682 & 96.29 & 90.66 & 93.97 \\ 
            & PDRCNN, XGBoost, CURNN & or & 93.23 & 93.31 & 0.1483 & 90.13 & 96.49 & 93.32 \\ 
            & PDRCNN, FFNN & majority & 93.82 & 93.23 & 18.7363 & 97.17 & 89.28 & 93.72 \\ 
            & eXpose, XGBoost & or & 93.04 & 93.17 & 0.0404 & 89.73 & 96.61 & 93.13 \\ 
            & CURNN, FFNN & majority & 93.64 & 93.04 & 18.7053 & 96.86 & 89.22 & 93.54 \\ 
            & eXpose, PDRCNN, XGBoost & or & 92.77 & 92.98 & 0.1110 & 89.24 & 96.71 & 92.87 \\ 
            & eXpose, XGBoost, CURNN & or & 92.54 & 92.80 & 0.0802 & 88.88 & 96.72 & 92.65 \\ 
            & eXpose, PDRCNN, XGBoost, CURNN & or & 92.32 & 92.65 & 0.1508 & 88.52 & 96.77 & 92.44 \\ 
        %%TOREMOVE-XGBoost & none & 92.58 & 92.30 & 0.0379 & 95.51 & 89.09 & 92.69 \\ 
        %%TOREMOVE-FFNN & none & 90.96 & 90.96 & 17.9453 & 87.97 & 93.95 & 90.86 \\ 
            & eXpose, XGBoost & majority & 91.42 & 90.77 & 0.0404 & 96.77 & 84.77 & 91.27 \\ 
            & PDRCNN, FFNN & or & 89.56 & 90.57 & 18.7362 & 84.54 & 96.60 & 89.74 \\ 
            & PDRCNN, XGBoost & majority & 91.21 & 90.56 & 0.1087 & 97.34 & 83.78 & 91.03 \\ 
            & XGBoost, CURNN & majority & 91.13 & 90.47 & 0.0779 & 96.98 & 83.96 & 90.96 \\ 
            & CURNN, FFNN & or & 89.36 & 90.47 & 18.7054 & 84.28 & 96.65 & 89.54 \\ 
            & XGBoost, FFNN & or & 89.14 & 90.16 & 18.7035 & 84.39 & 95.93 & 89.32 \\ 
            & PDRCNN, CURNN, FFNN & or & 88.82 & 90.14 & 18.7760 & 83.45 & 96.82 & 89.03 \\ 
            & eXpose, FFNN & or & 88.65 & 90.02 & 18.6681 & 83.22 & 96.81 & 88.86 \\ 
            & PDRCNN, XGBoost, FFNN & or & 88.49 & 89.92 & 18.7741 & 82.98 & 96.85 & 88.70 \\ 
            & eXpose, PDRCNN, FFNN & or & 88.40 & 89.86 & 18.7387 & 82.86 & 96.86 & 88.62 \\ 
            & XGBoost, CURNN, FFNN & or & 88.38 & 89.84 & 18.7433 & 82.84 & 96.84 & 88.60 \\ 
            & eXpose, CURNN, FFNN & or & 88.19 & 89.72 & 18.7079 & 82.57 & 96.86 & 88.41 \\ 
            & eXpose, PDRCNN, CURNN, FFNN & or & 87.99 & 89.60 & 18.7785 & 82.30 & 96.89 & 88.21 \\ 
            & PDRCNN, XGBoost, CURNN, FFNN & or & 87.86 & 89.52 & 18.8139 & 82.12 & 96.91 & 88.09 \\ 
            & eXpose, XGBoost, FFNN & or & 87.69 & 89.41 & 18.7060 & 81.89 & 96.92 & 87.93 \\ 
            & eXpose, PDRCNN, XGBoost, FFNN & or & 87.46 & 89.26 & 18.7766 & 81.58 & 96.94 & 87.70 \\ 
            & eXpose, XGBoost, CURNN, FFNN & or & 87.30 & 89.16 & 18.7458 & 81.37 & 96.94 & 87.54 \\ 
            & XGBoost, FFNN & majority & 89.76 & 89.07 & 18.7034 & 96.75 & 81.38 & 89.56 \\ 
            & All Detectors Combined & or & 87.11 & 89.04 & 18.8164 & 81.13 & 96.95 & 87.36 \\ 
        \bottomrule
	\end{tabular}
\end{table*}

\clearpage
\thispagestyle{empty}

\begin{table*}[htp!]
    
    \fontsize{6}{8}\selectfont
    % \small
    \centering
	\caption{Full description of methods' results on Wang's dataset (ordered by F1 score).
    }
	\label{tbls:tbl_detectors_scores_dataset2}
    \renewcommand{\arraystretch}{1.1}
	\setlength\tabcolsep{2.7pt}
	\begin{tabular}{clccccccc}
		\toprule
    	\multirow{2}{*}{} & Combination & Aggregation & AUC & F1 & Time & Precision & Recall & Accuracy\\
		& & & (\%) & (\%) & (sec) & (\%) & (\%) & (\%)\\
		\midrule
          & \textbf{\MethodName\_2} & DRL & 97.66 & 97.36 & 29.993 & 97.98 & 96.74 & 98.27 \\ 
          & SPIREL\_2 & DRL & 96.99 & 97.01 & 30.035 & 98.02 & 96.00 & 97.11  \\
          & \MethodName\_1 & DRL & 97.32 & 97.00 & 27.1499 & 99.60 & 94.40 & 98.23 \\ 
             & All Detectors Combined & stacking(RF) & 97.42 & 96.96 & 31.041 & 98.02 & 95.92 & 98.32 \\ 
          & SPIREL\_1 & DRL & 96.94 & 96.75 & 28.074 & 98.79 & 94.71 & 97.09\\ 
             & All Detectors Combined & boosting(ADB) & 96.93 & 96.71 & 31.247 & 96.26 & 97.16 & 97.11 \\ 
          & \textbf{\MethodName\_3} & DRL & 96.91 & 96.71 & 7.118 & 99.34 & 94.08 & 97.54 \\ 
          & SPIREL\_3 & DRL & 96.73 & 96.46 & 7.216 & 98.92 & 93.99 & 97.02 \\
             & All Detectors Combined & stacking(DT) & 97.22 & 96.44 & 31.0402 & 97.53 & 95.38 & 98.03 \\ 
          & \MethodName\_4 & DRL & 96.38 & 96.24 & 3.886 & 99.50 & 92.99 & 97.55 \\ 
          & SPIREL\_4 & DRL & 95.96 & 95.21 & 3.530 & 99.08 & 91.34 & 96.28  \\
          & \MethodName\_5 & DRL & 95.65 & 94.60 & 0.0075 & 95.81 & 93.39 & 96.21 \\ 
          & SPIREL\_5 & DRL & 95.59 & 94.55 & 0.008 & 95.60 & 93.50 & 96.03 \\ 
        %%TOREMOVE-CURNN & none & 95.62 & 94.52 & 0.01831 & 95.40 & 93.63 & 96.23 \\ 
             & eXpose, XGBoost, CURNN, FFNN & majority & 94.97 & 94.22 & 30.99175 & 98.13 & 90.62 & 96.88 \\ 
             & eXpose, PDRCNN, CURNN, FFNN & majority & 95.06 & 94.15 & 30.9965 & 97.48 & 91.04 & 96.82 \\ 
             & eXpose, PDRCNN, XGBoost, FFNN & majority & 94.83 & 94.02 & 31.0143 & 97.91 & 90.42 & 96.76 \\ 
        %%TOREMOVE-eXpose & none & 95.56 & 93.94 & 0.0024 & 92.94 & 94.95 & 95.75 \\ 
             & PDRCNN, XGBoost, CURNN, FFNN & majority & 94.53 & 93.75 & 31.0317 & 98.13 & 89.74 & 96.63 \\ 
        %%TOREMOVE-PDRCNN & none & 94.78 & 93.17 & 0.039 & 93.14 & 93.20 & 95.26 \\ 
             & eXpose, CURNN, FFNN & majority & 94.75 & 92.74 & 30.9561 & 93.43 & 92.04 & 95.61 \\ 
             & XGBoost, CURNN, FFNN & majority & 94.37 & 92.59 & 30.9892 & 94.19 & 90.98 & 95.54 \\ 
             & PDRCNN, CURNN, FFNN & majority & 94.29 & 92.36 & 30.9948 & 93.73 & 90.99 & 95.41 \\ 
             & eXpose, PDRCNN, FFNN & majority & 94.49 & 92.24 & 30.9789 & 92.62 & 91.85 & 95.33 \\ 
             & PDRCNN, XGBoost, FFNN & majority & 94.18 & 92.20 & 31.0133 & 93.53 & 90.86 & 95.32 \\ 
             & eXpose, XGBoost, FFNN & majority & 95.85 & 92.08 & 30.9737 & 92.24 & 91.92 & 94.51 \\ 
             & eXpose, FFNN & majority & 89.04 & 91.90 & 30.9368 & 96.20 & 87.60 & 92.90 \\ 
             & eXpose, XGBoost, CURNN & majority & 93.83 & 91.51 & 0.0586 & 92.45 & 90.56 & 94.93 \\ 
             & eXpose, PDRCNN, XGBoost, CURNN & majority & 93.25 & 91.47 & 0.1001 & 94.25 & 88.68 & 94.93 \\ 
             & PDRCNN, XGBoost, CURNN & majority & 93.42 & 91.11 & 0.0985 & 92.48 & 89.73 & 94.72 \\ 
             & eXpose, PDRCNN, XGBoost & majority & 93.63 & 91.08 & 0.0832 & 91.69 & 90.47 & 94.69 \\ 
             & eXpose, PDRCNN, CURNN & majority & 93.59 & 91.08 & 0.0638 & 91.82 & 90.34 & 94.69 \\ 
             & eXpose, CURNN & majority & 92.75 & 90.96 & 0.02071 & 94.27 & 87.65 & 94.66 \\ 
        %%TOREMOVE-XGBoost & none & 92.54 & 90.65 & 0.0377 & 92.23 & 89.06 & 93.61 \\ 
             & PDRCNN, CURNN & majority & 92.11 & 90.40 & 0.0614 & 94.52 & 86.28 & 94.35 \\ 
             & eXpose, PDRCNN & majority & 92.33 & 90.32 & 0.04451 & 93.54 & 87.09 & 94.31 \\ 
             & eXpose, CURNN & or & 93.43 & 89.22 & 0.0207 & 85.61 & 92.83 & 93.37 \\ 
             & PDRCNN, CURNN & or & 93.14 & 89.12 & 0.0604 & 86.41 & 91.82 & 93.39 \\ 
             & XGBoost, CURNN & or & 93.28 & 89.04 & 0.056 & 85.54 & 92.54 & 93.28 \\ 
             & eXpose, XGBoost & majority & 89.99 & 88.34 & 0.0402 & 94.81 & 81.87 & 93.22 \\ 
             & eXpose, PDRCNN & or & 92.96 & 88.29 & 0.0445 & 83.84 & 92.74 & 92.73 \\ 
             & XGBoost, CURNN & majority & 89.66 & 88.07 & 0.0556 & 95.01 & 81.13 & 93.07 \\ 
             & PDRCNN, XGBoost & or & 92.79 & 88.06 & 0.0798 & 83.59 & 92.53 & 92.59 \\ 
             & eXpose, PDRCNN, CURNN & or & 92.97 & 87.97 & 0.0628 & 82.42 & 93.52 & 92.40 \\ 
             & eXpose, XGBoost & or & 92.98 & 87.88 & 0.0401 & 82.47 & 93.29 & 92.42 \\ 
             & PDRCNN, XGBoost, CURNN & or & 92.89 & 87.81 & 0.0981 & 82.04 & 93.58 & 92.27 \\ 
             & eXpose, XGBoost, CURNN & or & 92.93 & 87.66 & 0.0584 & 81.15 & 94.17 & 92.07 \\ 
             & PDRCNN, XGBoost & majority & 89.25 & 87.50 & 0.07998 & 94.49 & 80.50 & 92.77 \\ 
             & CURNN, FFNN & majority & 88.40 & 87.37 & 30.9519 & 96.71 & 78.03 & 92.61 \\ 
             & eXpose, PDRCNN, XGBoost & or & 92.55 & 86.97 & 0.0822 & 79.77 & 94.16 & 91.53 \\ 
             & PDRCNN, FFNN & majority & 88.07 & 86.92 & 30.9752 & 96.36 & 77.48 & 92.38 \\ 
             & CURNN, FFNN & or & 92.94 & 86.87 & 30.9517 & 76.84 & 96.90 & 90.89 \\ 
             & eXpose, PDRCNN, XGBoost, CURNN & or & 92.45 & 86.67 & 0.1005 & 78.82 & 94.52 & 91.22 \\ 
        %%TOREMOVE-FFNN & none & 89.65 & 86.29 & 29.1686 & 85.24 & 87.34 & 90.36 \\ 
             & eXpose, FFNN & or & 92.33 & 85.85 & 30.9358 & 74.53 & 97.17 & 89.90 \\
             & XGBoost, FFNN & or & 92.05 & 85.57 & 30.9711 & 74.80 & 96.33 & 89.86 \\ 
             & PDRCNN, FFNN & or & 92.38 & 85.47 & 30.9755 & 74.14 & 96.79 & 90.13 \\ 
             & PDRCNN, CURNN, FFNN & or & 92.09 & 85.46 & 30.9938 & 73.65 & 97.26 & 89.50 \\ 
             & eXpose, CURNN, FFNN & or & 91.98 & 85.27 & 30.9541 & 73.11 & 97.43 & 89.27 \\ 
             & XGBoost, CURNN, FFNN & or & 91.95 & 85.23 & 30.9894 & 73.07 & 97.39 & 89.24 \\ 
             & XGBoost, FFNN & majority & 86.08 & 84.90 & 30.9721 & 96.27 & 73.53 & 91.26 \\ 
             & eXpose, PDRCNN, FFNN & or & 91.59 & 84.71 & 30.9779 & 71.99 & 97.43 & 88.72 \\ 
             & PDRCNN, XGBoost, FFNN & or & 91.52 & 84.61 & 31.0132 & 71.83 & 97.39 & 88.63 \\ 
             & eXpose, XGBoost, FFNN & or & 91.32 & 84.31 & 30.9735 & 71.07 & 97.55 & 88.27 \\ 
             & eXpose, PDRCNN, CURNN, FFNN & or & 91.31 & 84.29 & 30.9962 & 71.03 & 97.55 & 88.25 \\ 
             & PDRCNN, XGBoost, CURNN, FFNN & or & 91.21 & 84.16 & 31.0315 & 70.77 & 97.55 & 88.12 \\ 
             & eXpose, XGBoost, CURNN, FFNN & or & 91.04 & 83.91 & 30.9918 & 70.16 & 97.66 & 87.81 \\ 
             & eXpose, PDRCNN, XGBoost, FFNN & or & 90.70 & 83.46 & 31.0156 & 69.25 & 97.66 & 87.33 \\ 
             & All Detectors Combined & or & 90.47 & 83.16 & 31.0339 & 68.59 & 97.72 & 86.98 \\ 
        \bottomrule
        
	\end{tabular}
\end{table*}

\clearpage

\section{Transfer Theorem and Algorithm}
\label{sec:Appendix_transfer_theorem_algo}

\subsection{Density Function-based Security Transfer}
\label{subsec:Appendix_transfer_theorem}

Let $K \in \mathbb{N} \setminus \{1\}$ and  $\{d_n^k\}_{k=0}^K \subseteq \mathbb{R}$ be a strictly increase sequence, such that $T_n=[d_n^0,d_n^K]$ is an interval in $\mathbb{R}$. In our method, $d_n^K$ is the $s$-percentile of the dataset $D_n$, where $s \in \left(0,1\right]$. 

Let $\{ L_n^i \}_{i=1}^K$ be a partition of $T_n$, such that $\forall i=1,2,...,K$, each $L_n^i = \left[d_n^{i-1}, d_n^i\right]$ represents the $i$-th interval of the $i$-th case monotonic-increasing function $\{C_n^i (t)\}_{i=1}^K$. Thus, the function $C_n(t)$ is defined as follows:
\begin{equation}
    C_n(t) = \text{min} \left\{ \sum_{i=1}^K \textbf{I}_{L_n^i}(t) C_n^i(t), u \right\} 
    \label{funcs:source_domain_n_cases}
\end{equation}
where $\forall i=1,2,...,K$, $C_n^i(t)$ is elementary and  monotonic-increasing in $L_n^i$, and $\textbf{I}_{B}(t)$ is the indicator function of the set $B \subseteq \mathbb{R}$, which is defined as follows:
\begin{equation}
\textbf{I}_{B}(t) =
    \begin{cases}
      1, & \text{$t \in B$}\\
      0, &   \text{$t \notin B$}\\
    \end{cases}
\label{funcs:indicator_I_L_K}
\end{equation}

Note, we require that $\forall m=1,2,...,K-1$, $C_n^m\left(d_n^m\right)=C_n^{m+1}\left(d_n^m\right)$, and also define a constant $u = C_n^K \left(d_n^K\right)$ to ensure the continuity of $C_n(t)$. 

For a source domain $n=n_1$ with dataset $D_{n_1}$ and partitions $\{ L_{n_1}^i \}_{i=1}^K$, and for a destination domain $n={n_2} \ne {n_1}$ with dataset $D_{n_2}$ and partitions $\{L_{n_2}^i\}_{i=1}^K$, we require a proportional difference between the derivatives of each $i$-th case function, such that $C_{n_2}^i(t) \in \Theta \left( C_{n_1}^i(t) \right)$.

Next, we define a linear function $g:L_{n_2}^{i} \rightarrow L_{n_1}^{i}$ as follows:

\begin{equation}
    g_{{n_2}\rightarrow{n_1}}^i(t) = d_{n_1}^{i-1} + \frac{\left( d_{n_1}^i - d_{n_1}^{i-1} \right) \left(t - d_{n_2}^{i-1} \right)}{d_{n_2}^i - d_{n_2}^{i-1}} 
\end{equation}

Such that, $\forall i=1,2,...,K$, $C_{n_2}^i(t)$ is a composition of $C_{n_1}^i(t)$ and linear function $g_{{n_2}\rightarrow{n_1}}^i(t)$ as follows:
\begin{equation}
    C_{n_2}^i(t) = C_{n_1}^i \left( g_{{n_2}\rightarrow{n_1}}^i(t) \right)
\end{equation}

Next, we define a continuous random variable $X_n$ with the probability density function (i.e., PDF) as follows:
\begin{equation}
f_{X_n}(t) =
    \begin{cases}
      \frac{C_{n}(t)}{k_n} & ,\text{$ t \in T_n$}\\
      0 &   ,\text{otherwise}\\
    \end{cases}
\label{funcs:PDF_Xn_func_ext}
\end{equation}
\noindent where $k_n$ is defined as follows:
\begin{equation}
    k_n = \int_{T_n} C_n(t) dt = \sum_{i=1}^K \int_{L_n^i} C_n^i(t) dt
    \label{funcs:PDF_kn_norm_ext}
\end{equation}
Equation~\ref{funcs:PDF_kn_norm_ext}, enables us to ensure that the properties of $f_{X_n}(t)$ are those of a PDF. 

Next, we define a discrete random variable $Y_n$ as follows:
\begin{equation}
Y_n = \sum_{i=1}^{K} (i-1) \textbf{I}_{L_n^{i}} (X_n) = \sum_{i=2}^{K} (i-1) \textbf{I}_{L_n^{i}} (X_n) 
\label{funcs:indicator_Yn_func_ext}
\end{equation}
Such that $Y_n \in \{0,1,...,K-1\}$.

Because $Y_n$ is a discrete random variable, we can create a vectorized representation of the probability mass function $P_{Y_n} \in [0,1]^K$, where the $i$-th entry is defined as follows:
\begin{equation}
\mathbb{P}(Y_n = i-1) = \int_{L_n^{i}} f_{X_n}(t) dt
\label{funcs:probability_Yn_i_ext}
\end{equation}
\noindent where $i = 1,2,...,K$.

Note, from equations \ref{funcs:PDF_Xn_func_ext}-\ref{funcs:probability_Yn_i_ext}, we derive the following:
\begin{equation}
\begin{aligned}
& \sum_{i=1}^{K} \mathbb{P}(Y_n = i-1)  = \sum_{i=1}^{K} \int_{L_n^{i}} f_{X_n}(t) dt \\ 
& = \int_{\cup_{i=1}^{K}L_n^{i}} f_{X_n}(t) dt = \int_{T_n} f_{X_n}(t) dt=1
\end{aligned}
\label{funcs:prob_sum_to_one}
\end{equation}
\noindent where $i = 1,2,...,K$.

Based on equation~\ref{funcs:probability_Yn_i_ext}, we define the likelihood ratios' vector $\beta_n = (\beta_n^1,\beta_n^2,...,\beta_n^{K})^T \in \mathbb{R}^{K}$, where the $i$-th entry is defined as follows:

\begin{equation}
\beta_n^i = \frac{\mathbb{P}(Y_n = i-1)}{\mathbb{P}(Y_n = 0)} = \frac{\int_{L_n^{i}} f_{X_n}(t)dt}{\int_{{L_n^1}} f_{X_n}(t)dt} = \frac{\int_{L_n^{i}} C_n^{i}(t)dt}{\int_{L_n^1} C_n^1(t)dt}
\label{funcs:betas_Yn_i_ext}
\end{equation}
\noindent where $i = 1,2,...,K$.

Based on equations~\ref{funcs:prob_sum_to_one} and \ref{funcs:betas_Yn_i_ext}, we derive the following:

\begin{equation}
\mathbb{P}(Y_n = i-1)=\frac{\beta_n^i}{1_{K}^T\beta_n} 
\label{equations:bn_y1_y0_ext}
\end{equation}

\noindent where $i=1,2,...,K$ and $1_{K} \in \mathbb{R}^{K}$ is a column vector, which all his entries are 1.

For the datasets $D_{n_1}$ and $D_{n_2}$, the Kullback–Leibler (i.e., KL) divergence \cite{raiber2017kullback} between $Y_{n_1}$ and $Y_{n_2}$ is defined as follows:

\begin{equation}
D_{KL} \left( Y_{n_1} \| Y_{n_2} \right) = \sum_{y=0}^{K-1} \mathbb{P}(Y_{n_1}=y) \log_{10} \left( \frac{\mathbb{P}(Y_{n_1}=y)}{\mathbb{P}(Y_{n_2}=y)} \right)
\label{equations:D_KL_i_j_n}
\end{equation}

To require that the density of $D_{n_2}$ is similar to $D_{n_1}$, and since the KL metric is not symmetrical~\cite{tabibian2015speech}, we require that:
\begin{equation}
0 <
\frac{D_{KL} \left( Y_{n_1} \| Y_{n_2} \right)+D_{KL} \left( Y_{n_2} \| Y_{n_1} \right)}{2} < \varepsilon
\label{inequalities:avg_yi_yj_n}
\end{equation}
\noindent where $\varepsilon$ is hyper-parameter tolerance that decide how much the distributions could differ.
Since the KL measure is a convex function\cite{markechova2017kullback}, and its global minimum is $0$ (i.e., where distributions $Y_{n_1}, Y_{n_2}$ are the same, by Gibbs' inequality\cite{baccelli2012palm}), we require that the inequality would be greater than $0$. This requirement is derived from the assumption that the distributions $Y_{n_1}, Y_{n_2}$ are not the same (i.e., the domains are in fact different). Note, for destination dataset $n={n_2}$, the function $C_{n_2}$ depends on $K+1$ variables: $\{d_{n_2}^k\}_{k=0}^K$, and the only $K-1$ variables we could find are $\{\beta_{n_2}^k\}_{i=2}^{K-1}$ (i.e., $\beta_{n_2}^1$ is always 1), we require that $T_{n_2} = \left[d_{n_2}^0,d_{n_2}^K\right] \subseteq \mathbb{R}$ is known (i.e., $d_{n_2}^0, d_{n_2}^K$ are known). The algorithm to the presented above is shown below.
%in Appendix~\ref{subsec:Appendix_transfer_algo}.
\clearpage

%\subsection{Density Function-based Security Transfer Algorithm}
\label{subsec:Appendix_transfer_algo}
\begin{algorithm}[h]
 \caption{Density Function-based Security Transfer From S to D}
 \begin{algorithmic}[1]
 \STATE \textbf{input}:
 \\ $K \in \mathbb{N} \setminus \{1\}$
 \\ $\{d_S^k\}_{k=0}^K \in \mathbb{R}$ ; strictly increase sequence.
 \\ $d_D^0, d_D^k$ ; subset -- strictly increase sequence  $\{d_D^k\}_{k=0}^K \in \mathbb{R}$.
 \\ $d_S^K, d_D^K$ ; s-percentiles of datasets $S, D$, respectively.
 \\ $T_S = \left[d_S^{0},d_S^K\right]$, $T_D = \left[d_D^{0},d_D^K\right]$.
  \\ $\forall i=1,2,...,K,$ monotonic elementary function $C_S^i$:
  \\ $C_S^i:\left[d_S^{i-1},d_S^i\right] \rightarrow \mathbb{R}_{\geq 0}, \left(C_S^i\right)^\prime :\left[d_S^{i-1},d_S^i\right] \rightarrow \mathbb{R}_{\geq 0}$.
    \\ $\forall m=1,2,...,K-1,\;$ $C_S^m\left(d_S^m\right)=C_S^{m+1}\left(d_S^m\right)$.
    \\ $\varepsilon\ >0 $ ; distribution tolerance rate.
    \\ $\alpha > 0$ ; learning rate.

  \STATE \textbf{for} $i=1,2,...,K$
  \\ $L_S^k = \left[d_S^{k-1},d_S^k\right]$
  \\ $L_D^k = \left[d_D^{k-1},d_D^k\right]$
  \\ $C_D^i\left(t;d_D^{i-1}, d_D^i\right) = C_S\left(  d_{S}^{i-1} + \frac{\left( d_{S}^i - d_{S}^{i-1} \right) \left(t - d_{D}^{i-1} \right)}{d_{D}^i - d_{D}^{i-1}}  \right)$ 
  \\ $\beta_S^i= \int_{L_S^{i}} C_S^{i}(t)dt \left( \int_{L_S^1} C_S^1(t)dt \right)^{-1} $
  \\ \textbf{end for}
  \STATE \textbf{define}:
  \\ $u_S = C_S^K \left(d_S^K\right)$, $u_D = C_D^K \left(d_D^K\right)$
  \\ $C_S(t) = \text{min} \left\{ \sum_{i=1}^K \textbf{I}_{L_S^i}(t) C_S^i(t), u_S \right\}$
  \\ $C_D(t) = \text{min} \left\{ \sum_{i=1}^K \textbf{I}_{L_D^i}(t) C_D^i(t), u_D \right\}$
  \\ $\beta_S = (1,\beta_S^2,...,\beta_S^{K})^T \in \mathbb{R}_{\geq 0}^K$ 
  \\ Sample $\theta=(\theta^1,\theta^2,...,\theta^{K-1})^T \in \mathbb{R}_{\geq 0}^{K-1}$ randomly.
  \\ $\psi \;=(1,\theta^1,\theta^2,...,\theta^{K-1})^T = \left(1, \theta^T\right)^T \in \mathbb{R}_{\geq 0}^{K}$
  \\ Random vars. : $Y_S,Y_D \in \{0,1,...,K-1\} $ w.p.:
  \\$P_{Y_S} = \left( 1_K^T \beta_S \right)^{-1} \beta_S , P_{Y_D} = \left(1_K^T \psi \right)^{-1} \psi$
  \\ $h(\psi; \beta_S) = \frac{1}{2} \left( D_{KL} \left( Y_S \| Y_D \right) +  D_{KL} \left( Y_D \| Y_S \right) \right)$
  %\STATE \textbf{while} $h(\psi; \beta_S) \leq \delta_1$ 
  %\\ Sample $\theta$ randomly from $\in \mathbb{R}_{\geq 0}^{K-1}$.
  %\\ $\psi \gets (1,\theta^T)$ 
  \STATE \textbf{while} $h(\psi; \beta_S) \geq \varepsilon$
  \\ $\theta \gets \theta - \alpha \nabla_{\theta}
  h(\psi ; \beta_S)$
  \\ $\psi \gets \left(1, \theta_T \right)^T$
  \\ \textbf{end while}
 \STATE $\beta_D = \psi$
 \STATE \textbf{for} $m=1,...,K-1$
    \\ derive $d_D^{m}$ from the equation:
    \\ $\sum_{i=m+1}^K \beta_D^i = \int_{d_D^m}^{d_D^K} C_D(t)dt \left( \int_{L_D^1} C_D^1(t)dt \right)^{-1} $
  \\ \textbf{end for}
 \STATE \textbf{return} $C_D(t)$ 
 \end{algorithmic} 
 \end{algorithm}

\end{document}